\documentclass[sigplan,nonacm]{acmart}
\usepackage[normalem]{ulem}

\usepackage{color}
\usepackage{listings}
\usepackage[]{hyperref}
\usepackage{amsmath,amsfonts}
\usepackage{mathtools}
\usepackage[ruled, vlined, linesnumbered]{algorithm2e}
\usepackage{circledsteps}
\usepackage{comment}
\usepackage{booktabs}
\usepackage{wasysym}
\usepackage{multirow}
\usepackage{graphicx}
\usepackage{makecell}
\usepackage{subcaption}
\usepackage{xcolor}
\usepackage{mathpartir}
\usepackage{adjustbox}
\usepackage{mdframed}
\usepackage{siunitx}
\usepackage{pifont}% http://ctan.org/pkg/pifont
\newcommand{\gcmark}{\textcolor{green}{\textbf{\ding{51}}}}
\newcommand{\rxmark}{\textcolor{red}{\textbf{\ding{55}}}}

\usepackage{wasysym}

\usepackage[frozencache=true,cachedir=minted-cache]{minted} 
\usepackage{tikz}
\usetikzlibrary{tikzmark,fit,backgrounds}

\newcommand{\Name}{Dato\xspace}

\newcommand{\Stream}[2]{\mathsf{Stream}[#1,#2]}
\newcommand{\LFut}[1]{\mathsf{LFuture}\,#1}
\newcommand{\Free}{\mathsf{Free}}
\newcommand{\Used}{\mathsf{Used}}
\newcommand{\sid}{\mathsf{id}}

\newcommand{\showcomments}{yes}

\newcommand\fixme[1]{
    \ifthenelse{\equal{\showcomments}{yes}}{\textcolor{red}{#1}}{\ignorespaces}
}
\newcommand\hz[1]{
    \ifthenelse{\equal{\showcomments}{yes}}{\textcolor{red}{\small [hz: #1~]}}{\ignorespaces}
}
\newcommand\sh[1]{
    \ifthenelse{\equal{\showcomments}{yes}}{\textcolor{purple}{\small [sh: #1~]}}{\ignorespaces}
}
\newcommand\zz[1]{
    \ifthenelse{\equal{\showcomments}{yes}}{\textcolor{blue}{\small [zz: #1]}}{\ignorespaces}
}

\newcommand\ns[1]{
    \ifthenelse{\equal{\showcomments}{yes}}{\textcolor{cyan}{\small [ns: #1~]}}{\ignorespaces}
}

\begin{document}
\title{\Name: A Task-Based Programming Model\\for Dataflow Accelerators}

% \author{Shihan Fang}
% \authornote{Equal Contribution.}
% \authornote{Visiting student from Shanghai Jiao Tong University.}
% \affiliation{%
%   \institution{Shanghai Jiao Tong University}
%   \city{Shanghai}
%   \state{}
%   \country{China}
% }
% \email{fang-account@sjtu.edu.cn}
% \author{Hongzheng Chen}
% \authornotemark[1]
% \affiliation{%
%   \institution{Cornell University}
%   \city{Ithaca}
%   \state{New York}
%   \country{USA}
% }
% \email{hzchen@cs.cornell.edu}
% \author{Niansong Zhang}
% \affiliation{%
%   \institution{Cornell University}
%   \city{Ithaca}
%   \state{New York}
%   \country{USA}
% }
% \email{nz264@cornell.edu}
% \author{Jiajie Li}
% \affiliation{%
%   \institution{Cornell University}
%   \city{Ithaca}
%   \state{New York}
%   \country{USA}
% }
% \email{jl4257@cornell.edu}
% \author{Han Meng}
% \authornote{Work done while at Cornell.}
% \affiliation{%
%   \institution{UC Merced}
%   \city{Merced}
%   \state{California}
%   \country{USA}
% }
% \email{hanmeng@ucmerced.edu}
% \author{Adrian Liu}
% \authornotemark[3]
% \affiliation{%
%   \institution{UCLA}
%   \city{Los Angeles}
%   \state{California}
%   \country{USA}
% }
% \email{adrianliu@ucla.edu}
% \author{Zhiru Zhang}
% \affiliation{%
%   \institution{Cornell University}
%   \city{Ithaca}
%   \state{New York}
%   \country{USA}
% }
% \email{zhiruz@cornell.edu}

\author{Shihan Fang\footnotemark[1]\footnotemark[2]\enspace
Hongzheng Chen\footnotemark[1]\enspace
Niansong Zhang\enspace
Jiajie Li\enspace\\
Han Meng\footnotemark[4]\footnotemark[3]\enspace
Adrian Liu\footnotemark[5]\footnotemark[3]\enspace
Zhiru Zhang\smallskip\\
\normalsize Cornell University\enspace
\footnotemark[4]UC Merced\enspace \footnotemark[5]UCLA\\
\href{mailto:hzchen@cs.cornell.edu}{\textsf{hzchen@cs.cornell.edu}},\enspace \href{mailto:zhiruz@cornell.edu}{\textsf{zhiruz@cornell.edu}}
}
\renewcommand{\shortauthors}{}

% add the paper content here
\begin{abstract}
Recent deep learning workloads increasingly push computational demand beyond what current memory systems can sustain, with many kernels stalling on data movement rather than computation. While modern dataflow accelerators incorporate on-chip streaming to mitigate off-chip bandwidth limitations, existing programming models struggle to harness these capabilities effectively. Low-level interfaces provide fine-grained control but impose significant development overhead, whereas high-level tile-based languages abstract away communication details, restricting optimization and forcing compilers to reconstruct the intended dataflow.

We present \Name, a Python-embedded, task-based programming model for dataflow accelerators that elevates data communication and sharding to first-class type constructs. Developers write programs as a graph of tasks connected via explicit stream types, with sharded inputs specified using layout types. These tasks are first mapped virtually onto the accelerator's spatial fabric, and the compiler then generates a physical mapping that respects hardware constraints.
Experimental results on both AMD Ryzen AI NPU and Alveo FPGA devices demonstrate that \Name achieves high performance while significantly reducing the burden of writing optimized code. On the NPU, \Name attains up to 84\% hardware utilization for GEMM and delivers a 2.81$\times$ speedup on attention kernels compared to a state-of-the-art commercial framework. On the FPGA, \Name surpasses leading frameworks in performance when generating custom systolic arrays, achieving 98\% of the theoretical peak performance.
\end{abstract}
% \zz{is this a standard term? can we use utilization or occupancy instead? I don't see it's formally defined anywhere in our paper}

\maketitle % should come after the abstract

% Footnotes must come after \maketitle to stay on this page:
\begingroup
\renewcommand{\thefootnote}{\fnsymbol{footnote}}
\footnotetext[1]{Equal Contribution.}
\footnotetext[2]{Visiting student from Shanghai Jiao Tong University.}
\footnotetext[3]{Work done at Cornell University.}
\endgroup

\section{Introduction}
\label{sec:intro}
\begin{table*}[t]
\caption{Comparison of existing accelerator programming frameworks.}
\label{tab:comparison}
\centering
\resizebox{0.82\linewidth}{!}{
\begin{tabular}{cccccccc}\Xhline{2\arrayrulewidth}
\textbf{Framework} & \textbf{Target} & \textbf{Input Language} & \thead{\textbf{Explicit}\\\textbf{Communication}} & \thead{\textbf{Explicit}\\\textbf{Sharding}} & \thead{\textbf{Automatic}\\\textbf{Optimization}} & \thead{\textbf{Type Checking}\\\textbf{on Dataflow}}\\\hline
CUTLASS~\cite{cutlass} & GPU & CUDA & \gcmark & \rxmark & \rxmark & \rxmark\\
Exo~\cite{yuka2022exo} & CPU/Xcel & Python & \rxmark & \rxmark & \rxmark  & \rxmark\\
Taskflow~\cite{huangtaskflow2022} &CPU/GPU & C++ & \gcmark & \rxmark & \gcmark & \rxmark \\
Triton~\cite{tillet2019triton}  & GPU & Python & \rxmark & \rxmark &\gcmark  & \rxmark\\\hline
Allo~\cite{chen2024allo}  & FPGA & Python/MLIR & \gcmark & \rxmark &\rxmark  & \rxmark\\
HIDA~\cite{ye2024hida} & FPGA & C++ & \rxmark  & \rxmark & \gcmark& \rxmark\\
TAPA~\cite{guo2023tapa} & FPGA & C++  & \gcmark & \rxmark & \gcmark & \rxmark\\\hline
ARIES~\cite{zhuang2025aries} & AIE/NPU & Python  & \rxmark & \rxmark & \gcmark & \rxmark\\
IRON~\cite{hunhoff2025iron} & NPU & Python/MLIR & \gcmark & \rxmark & \rxmark & \rxmark\\
\Name (Ours) & FPGA/NPU & Python & \gcmark & \gcmark & \gcmark & \gcmark\\\hline
\Xhline{2\arrayrulewidth}
\end{tabular}
}
\end{table*}

% \thead{\textbf{Explicit}\\\textbf{Dataflow}} & \thead{\textbf{Automatic}\\\textbf{Optimization}}
Large language models (LLMs) have driven computational demand to unprecedented levels, yet their performance is increasingly constrained by memory~\cite{chen2024understanding,pope2023scalellmtpu,vonneumann_bottleneck}.
While peak FLOPs and specialized tensor engines scale rapidly, the effective throughput for LLM training and inference is often bounded by how quickly the tensors can be moved, staged, and transformed.
Off-chip DRAM bandwidth improves only marginally and the energy cost per bit access far exceeds that of a FLOP, a limitation widely known as the \emph{memory wall}~\cite{amir2024memorywall}.
As a result, many kernels fail to reach their arithmetic limits, stalling on data movement, layout transformations, and synchronization.

One promising architectural response is to minimize off-chip traffic by exploiting on-chip dataflow: streaming data directly between hardware modules so that intermediate results avoid detours through DRAMs.
Dataflow accelerators such as Google TPU~\cite{jouppi2023tpuv4}, AWS Trainium~\cite{fu2024trainium}, and AMD Ryzen AI NPU~\cite{xdna_npu} embody this philosophy by coupling compute units with scratchpads and lightweight interconnects to sustain pipelines.
Even general-purpose GPUs have embraced similar principles: recent NVIDIA Hopper~\cite{hopper} and Blackwell~\cite{blackwell} GPUs introduce dedicated tensor memory accelerators (TMA) to stream tensors asynchronously, overlapping data transfers with computation. By making data movement a first-class concern through channels and FIFOs, these designs can transform memory-bound workloads into high-throughput pipelines that keep compute units saturated while greatly reducing off-chip traffic.

Despite these advances, current programming models struggle to fully unlock the potential of dataflow accelerators, which remain trapped in the longstanding trade-off between performance and productivity.

\textbf{Challenge 1: Low-level communication control improves performance but hinders productivity.} 
Low-level programming interfaces offer fine-grained hardware control but impose significant burdens on developers. For example, IRON~\cite{hunhoff2025iron}, a frontend for programming AMD NPUs, exposes all FIFO and DMA details to the user---similar in abstraction level to CUDA for GPUs. Even a simple matrix multiplication kernel can require hundreds of lines of code, as IRON essentially serves as a Python wrapper over the MLIR-AIE dialect~\cite{mliraie}.
Recent schedule languages such as Exo~\cite{yuka2022exo,ikarashi2025exo2} and Allo~\cite{chen2024allo} aim to decouple hardware customization from algorithm specification, enabling expressive transformations. However, in the dataflow setting, these approaches often exacerbate the problem. For instance, Allo provides a \texttt{.relay()} primitive to model communication, but developers still write array-based code and must manually translate it into FIFOs. This creates a semantic mismatch between array-oriented frontends and the inherently stream-oriented hardware backends. Complex producer-consumer topologies become verbose and brittle to express, while compilers must infer dataflow intent from array operations---adding complexity, risking correctness, and missing optimization opportunities. In effect, compilers are required to \emph{reconstruct} the dataflow a programmer intended, rather than enabling programmers to specify it directly.

\textbf{Challenge 2: Tile-based languages hide communication, limiting optimization.}
Tile-based programming models such as Triton~\cite{tillet2019triton} excel at expressing per-tile computation with implicit caching but offer little explicit control over inter-tile or inter-kernel communication. Efforts like ARIES~\cite{zhuang2025aries} adapt Triton-like syntax for dataflow accelerators, yet still enforce fixed communication patterns, restricting developers from fine-grained control. This limitation becomes especially acute in multi-kernel designs, where users have no choice to manage how data flows between kernels.
As a result, intermediate results are forced back to off-chip memory before being read again, missing the opportunity to exploit on-chip streaming and resulting in performance degradation.

We argue that \textbf{data communication must be a first-class abstraction} in programming models for dataflow accelerators. To this end, we propose \Name\footnote{A shorthand for \underline{D}ataflow \underline{a}cceleration with a \underline{t}ask-\underline{o}riented programming model. Code available at \href{https://github.com/cornell-zhang/allo}{https://github.com/cornell-zhang/allo}.}, a task-based programming model that elevates both data communication and sharding to first-class types.
% \zz{instead of calling out Allo, spell out the acronym, something like Dataflow Acceleration with a Task-Oriented Programming Model}
% \zz{we need to spell out the full name of ADF, since it's an acronym. Of course, we can also come up with alternative later}
In \Name, programmers define tasks connected by \emph{stream types} and pass in sharded data with \emph{layout types}.
These tasks are virtually onto the spatial fabric, with the compiler automatically translating virtual mappings into physical mappings. By co-designing compute and communication at the programming-model level, \Name explicitly exposes inter-task streaming, reduces off-chip traffic, and alleviates compiler development burdens.
In summary, our contributions are as follows:
% \zz{need to make contributions sound more compelling}
\begin{itemize}
\item We introduce \Name, a Python-embedded, task-based programming model that elevates data communication and layout to first-class, statically checked types, enabling developers to express multi-task pipelines with shard-aware dataflow directly.
\item We propose a virtual-to-physical mapping mechanism that lets users declare task placement, while the compiler automatically binds each task to appropriate processing engines, ensuring high performance at scale.
\item We demonstrate performance and portability on AMD Ryzen AI NPUs and an Alveo U280 FPGA. Across diverse benchmarks, \Name matches or surpasses hand-optimized baselines with substantially less code, achieving up to 84\% hardware utilization for GEMM and a 2.81$\times$ speedup on attention kernels over a commercial framework on the NPU, and generating a custom systolic array on FPGA that reaches 98\% of the theoretical peak throughput.
% \fixme{reaching 97\% of theoretical peak throughput for a systolic array running on the FPGA.} \zz{replace it with ``generating a custom systolic array on FPGA that reaches ... peak throughput.''}
\end{itemize}

\section{Background and Related Work}
\label{sec:background}
This section reviews representative dataflow accelerators along with their existing programming models.

\subsection{Dataflow Accelerators}
Systolic array architectures have become a cornerstone of modern deep learning acceleration due to their ability to sustain high throughput for dense linear algebra workloads. 
Architectures such as Google’s TPU~\cite{jouppi2017tpuv1,jouppi2023tpuv4,tpuv7}, AWS Trainium~\cite{fu2024trainium}, and Inferentia~\cite{aws_inferentia} use large grids of processing elements to stream matrix multiplications, reusing operands locally and passing partial results to neighbors to reduce off-chip memory traffic. Coupled with on-chip scratchpads and optimized memory controllers, they achieve high utilization and energy efficiency for GEMM- and convolution-heavy workloads.

Programmable dataflow fabrics extend the systolic philosophy to more diverse workloads by giving finer control over on-chip communication and computation. Cerebras WSE links hundreds of thousands of cores across a full wafer via a custom fabric for massive parallelism~\cite{cerebras_wse}, while AMD’s Ryzen AI NPU~\cite{xilinxAIE} employs an AI Engine array of VLIW processors and vector units connected through a programmable NoC to support both streaming and general-purpose tasks. Other accelerators use scalable meshes of AI cores with configurable memory and communication~\cite{emani2021sambanova,tenstorrent,ibm_spyre}. NVIDIA’s Hopper~\cite{hopper} and Blackwell~\cite{blackwell} architectures likewise add tensor memory accelerators (TMA) to stage and stream tensors asynchronously, overlapping data movement with computation for higher utilization.

\begin{figure}[t]
\centering
\includegraphics[width=\linewidth]{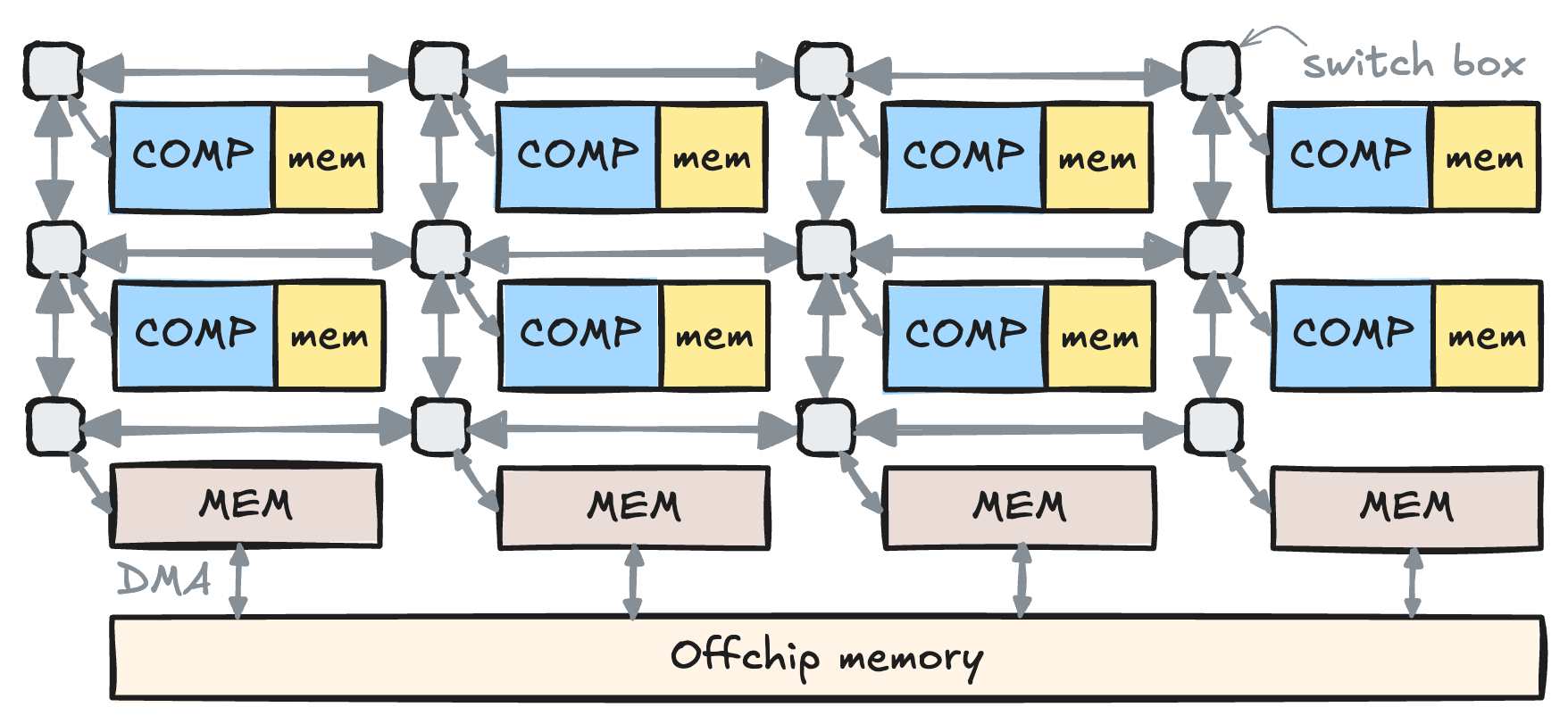}
\caption{A typical dataflow accelerator architecture~\cite{xilinxAIE,tenstorrent}. The actual number of hardware units may be greater.}
\label{fig:aie}
\Description{A typical dataflow accelerator.}
\vspace{-.1in}
\end{figure}

Fig.~\ref{fig:aie} illustrates a typical dataflow architecture: an array of compute units with local memory and streaming interfaces that exchange data with neighbors. On-chip switch boxes route traffic across the array, while scratchpad memories buffer tensors and support parallel access across tiles. Additional interface units connect to external memory and I/O via DMA engines. By explicitly orchestrating data movement, dataflow accelerators efficiently map workloads onto spatially distributed tasks connected by streams.

\subsection{Existing Programming Models}
Table~\ref{tab:comparison} compares \Name with other accelerator programming frameworks.

Early GPU programming relied heavily on CUDA, which offers fine-grained control over hardware and enables a wide range of performance optimizations. While this \textbf{low-level interface} is powerful, it requires significant expertise and manual effort. CUTLASS~\cite{cutlass} builds on CUDA with expert-tuned matrix kernels for deep learning workloads.
IRON~\cite{hunhoff2025iron} represents a similar effort for AMD NPUs: built atop the MLIR-AIE dialect~\cite{mliraie}, it offers built-in kernels for developers to write different applications. However, as accelerator architectures grow more complex, low-level abstractions increasingly hinder productivity, as developers must manage numerous hardware-specific details.

% To address this challenge, \textbf{schedule languages} such as Exo~\cite{yuka2022exo,ikarashi2025exo2} and Allo~\cite{chen2024allo} attempt to separate algorithm specification from hardware customization. Allo, for example, introduces the \texttt{.relay()} primitive to decouple communication from algorithm, enabling optimization of data movement independently from computation. However, this indirection can make it harder for users to reason about performance, particularly for dataflow accelerators where communication patterns are central to efficiency. In contrast, \Name elevates data communication to a first-class type, allowing programmers to define explicit dataflow connections directly in their code, thus improving both expressiveness and productivity.

% Another stream of work aims to boost productivity through \textbf{tile-based programming models}, which abstract away certain hardware details while exposing tile-level parallelism. Triton~\cite{tillet2019triton} exemplifies this approach for GPUs, offering a Python-based domain-specific language (DSL) that allows developers to write high-performance kernels without explicit thread-level programming. ARIES~\cite{zhuang2025aries} applies a similar tile-based abstraction to AIE~\cite{xilinxAIE} and NPUs, but its lack of fine-grained communication control can lead to suboptimal mappings and degraded performance for communication-intensive workloads. These approaches work well for compute-dominated kernels but are less suited to accelerators with complex inter-tile dataflows.

To address this challenge, schedule languages adopt the principle of separation of concerns by encapsulating optimization decisions as primitives~\cite{chen2018tvm,jrk2013halide,lai2019heterocl,xiang2022heteroflow,debjit2022dac,chen2024slapo}.
Exo~\cite{yuka2022exo,ikarashi2025exo2} and Allo~\cite{chen2024allo} are specifically designed for hardware customization.
In particular, Allo's \texttt{.relay()} primitive decouples communication from computation, enabling independent optimization of data movement. However, this indirection complicates performance reasoning, especially on dataflow accelerators where communication patterns dominate efficiency. 
In contrast, \Name treats data communication as a first-class type, letting programmers define explicit dataflow connections directly in code to improve expressiveness and productivity.

Another approach improves productivity through \textbf{tile-based programming models}, which abstract hardware details while exposing tile-level parallelism~\cite{tillet2019triton,hagedorn2023graphene,wang2025tilelang,ding2025tilus}. Triton~\cite{tillet2019triton} provides a Python-based DSL for high-performance GPU kernels without explicit thread programming, while ARIES~\cite{zhuang2025aries} applies a similar model to AIE~\cite{xilinxAIE} and NPUs. However, ARIES lacks fine-grained communication control, which prevents inter-kernel communication on-chip. 
Such models work well for compute-dominated kernels but are less effective for complex inter-tile dataflows.

Finally, \textbf{task-based programming models} express computation as interconnected tasks. Taskflow~\cite{huangtaskflow2022} enables in-graph control flow and heterogeneous CPU–GPU execution but leaves memory and device management to developers.
Several FPGA-oriented frameworks~\cite{ye2024hida,guo2023tapa,suhail2025streamhls} expose C++ interfaces that modularize designs via FIFOs, yet they require actual execution to uncover dataflow bugs, limiting static verification. For dataflow accelerators, we argue that tasks should be natively connected by streaming buffers in the programming model. \Name follows this direction with a Python-embedded, task-based interface where tasks communicate via explicit streams, and the compiler optimizes both mapping and communication, combining fine-grained control with high-level productivity.
In parallel, other works~\cite{dadu2022taskstream,rucker2024revet,ghosh2025ripple,li2025plaid} have proposed new domain-specific languages (DSLs) targeting non-commercial CGRAs or reconfigurable dataflow architectures, but these systems provide limited support for data layout management and dataflow type checking, which are essential for scalable and reliable accelerator programming.

\section{\Name Programming Model}
In this section, we first give an overview of \Name and discuss the details of the frontend programming interface.

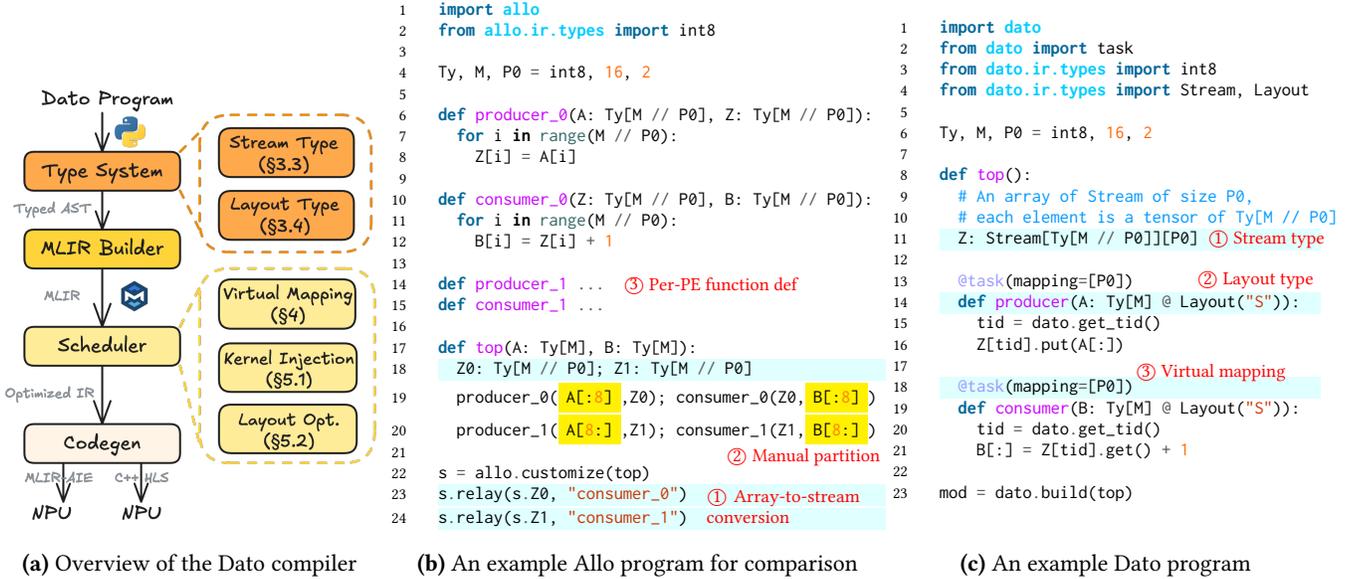
\begin{figure*}[t]
\begin{subfigure}[b]{0.29\linewidth}
\centering
\includegraphics[width=\linewidth]{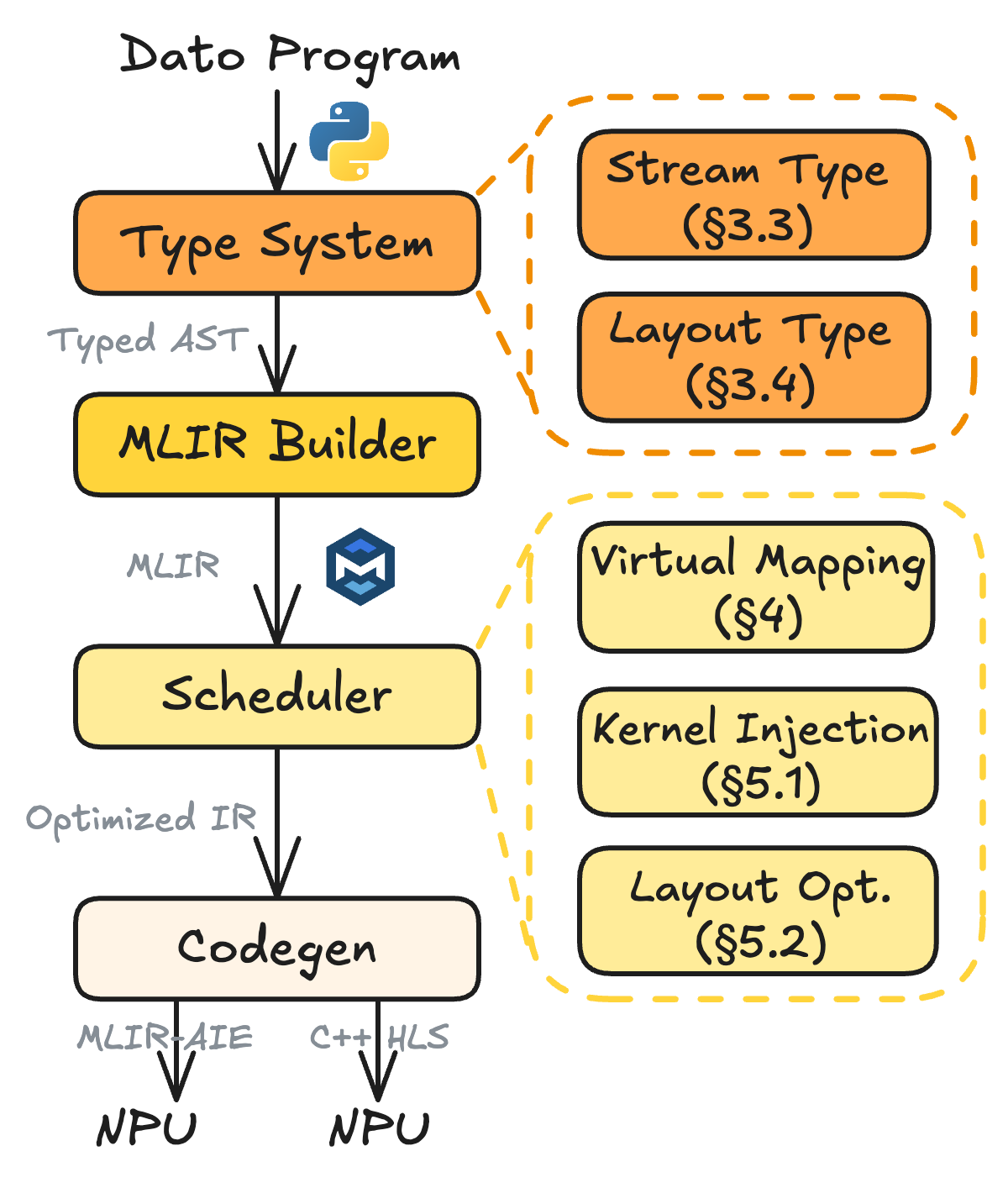}
\caption{Overview of the \Name compiler}
\label{subfig:overview}
\end{subfigure}
% \hfill\vline\hfill
\begin{subfigure}[b]{0.37\linewidth}
\begin{minted}[linenos,
               fontsize=\scriptsize,
               xleftmargin=1.8em,
               highlightlines={18,23,24},
               escapeinside=||,
               autogobble]{python}
import allo
from allo.ir.types import int8

Ty, M, P0 = int8, 16, 2

def producer_0(A: Ty[M // P0], Z: Ty[M // P0]):
  for i in range(M // P0):
    Z[i] = A[i]

def consumer_0(Z: Ty[M // P0], B: Ty[M // P0]):
  for i in range(M // P0):
    B[i] = Z[i] + 1

def producer_1 ... |\tikzmark{alloc}|
def consumer_1 ...

def top(A: Ty[M], B: Ty[M]):
  Z0: Ty[M // P0]; Z1: Ty[M // P0]
  producer_0(|\colorbox{yellow}{A[:\textcolor{orange}{8}]}|,Z0); consumer_0(Z0,|\colorbox{yellow}{B[:\textcolor{orange}{8}]}|)
  producer_1(|\colorbox{yellow}{A[\textcolor{orange}{8}:]}|,Z1); consumer_1(Z1,|\colorbox{yellow}{B[\textcolor{orange}{8}:]}||\tikzmark{allob}|)

s = allo.customize(top)
s.relay(s.Z0, "consumer_0") |\tikzmark{alloa}|
s.relay(s.Z1, "consumer_1")
\end{minted}
\caption{An example Allo program for comparison}
\label{subfig:allo}
\end{subfigure}
% \hfill\vline\hfill
\begin{subfigure}[b]{0.32\linewidth}
\begin{minted}[linenos,
               fontsize=\scriptsize,
               xleftmargin=1.8em,
               highlightlines={11,14,18},
               escapeinside=||]{python}
import dato
from dato import task
from dato.ir.types import int8
from dato.ir.types import Stream, Layout

Ty, M, P0 = int8, 16, 2

def top():
  # An array of Stream of size P0,
  # each element is a tensor of Ty[M // P0]
  Z: Stream[Ty[M // P0]][P0]|\tikzmark{stream}|

  @task(mapping=[P0])
  def producer(A: Ty[M] @ |\tikzmark{ls}|Layout("S")|\tikzmark{le}|):|\tikzmark{layout}|
    tid = dato.get_tid()
    Z[tid].put(A[:])

  @task(mapping=[P0])|\tikzmark{mapping}|
  def consumer(B: Ty[M] @ Layout("S")):
    tid = dato.get_tid()
    B[:] = Z[tid].get() + 1

mod = dato.build(top)
\end{minted}
\begin{tikzpicture}[overlay, remember picture]
  \node[anchor=west, font=\scriptsize, text=red]
    (label) at ([xshift=0cm,yshift=0.05cm]pic cs:stream) {\textcircled{1} Stream type};
  \node[anchor=west, font=\scriptsize, text=red]
    (label) at ([xshift=-1.5cm,yshift=0.35cm]pic cs:layout) {\textcircled{2} Layout type};
  \node[anchor=west, font=\scriptsize, text=red]
    (label) at ([xshift=-0.1cm,yshift=0.25cm]pic cs:mapping) {\textcircled{3} Virtual mapping};
  \node[anchor=west, font=\scriptsize, text=red, text width=2.5cm]
    (label) at ([xshift=0cm,yshift=-0.1cm]pic cs:alloa) {\textcircled{1} Array-to-stream conversion};
  \node[anchor=west, font=\scriptsize, text=red]
    (label) at ([xshift=-2cm,yshift=-0.3cm]pic cs:allob) {\textcircled{2} Manual partition};
  \node[anchor=west, font=\scriptsize, text=red]
    (label) at ([xshift=0cm,yshift=0.05cm]pic cs:alloc) {\textcircled{3} Per-PE function def};
  % \begin{scope}[on background layer] % draw *behind* text
  %  \node[fit=(pic cs:ls)(pic cs:le),
  %         fill=yellow, fill opacity=.5, draw=none] {};
  % \end{scope}
  % \draw[->, thick] (label.west) -- ([xshift=0.15cm]pic cs:imp);
\end{tikzpicture}
\caption{An example \Name program}
\label{subfig:dato}
\end{subfigure}
% \hfill
\caption{Example Allo and \Name programs of a simple producer-consumer pattern.}
\label{fig:overview}
% required by acmart, for voiceover to read the paper aloud
\Description{}
\end{figure*}

\subsection{Overview}
\label{sub:overview}
To maximize usability and productivity, \Name is designed as a Python-embedded accelerator programming language. It builds on top of the open-source accelerator programming framework Allo~\cite{chen2024allo}
and reuses the MLIR infrastructure~\cite{chris2021mlir}, while introducing significant new capabilities tailored to dataflow accelerator programming.
% \zz{we may want to hide Allo for anonymity}
As illustrated in Fig.~\ref{subfig:overview}, the compilation process begins with a Python function written by the user. This function first undergoes dataflow-specific type checking (\S~\ref{sub:stream} and \S~\ref{subsec:layout}), ensuring that communication patterns between tasks and data sharding are explicitly validated. The resulting type-annotated abstract syntax tree (AST) is then lowered through an MLIR builder to construct the intermediate representation (IR).

With the IR constructed, \Name generates a virtual computation graph that captures the program's dataflow structure. Because dataflow accelerators often feature a limited number of processing engines (PEs), while user programs may express significantly larger computations, \Name performs a virtual-to-physical mapping step (\S~\ref{sec:mapping}). This step bridges the gap between logical parallelism and finite hardware resources, enabling efficient task placement and scheduling across the spatial fabric.

Following physical mapping, the compiler applies kernel injection and layout optimization passes (\S~\ref{sec:opt}) to refine memory placement and compute operations. The final optimized MLIR module is then lowered to backend-specific code. To target AMD NPUs, we leverage MLIR-AIE~\cite{mliraie} as the backend; for FPGAs, we generate C++ code for high-level synthesis (HLS) and synthesize it using AMD Vitis~\cite{vitis_hls_2023}.

\subsection{A Motivating Example}
\label{sub:motivation}
\begin{figure}[t]
\centering
\includegraphics[width=\linewidth]{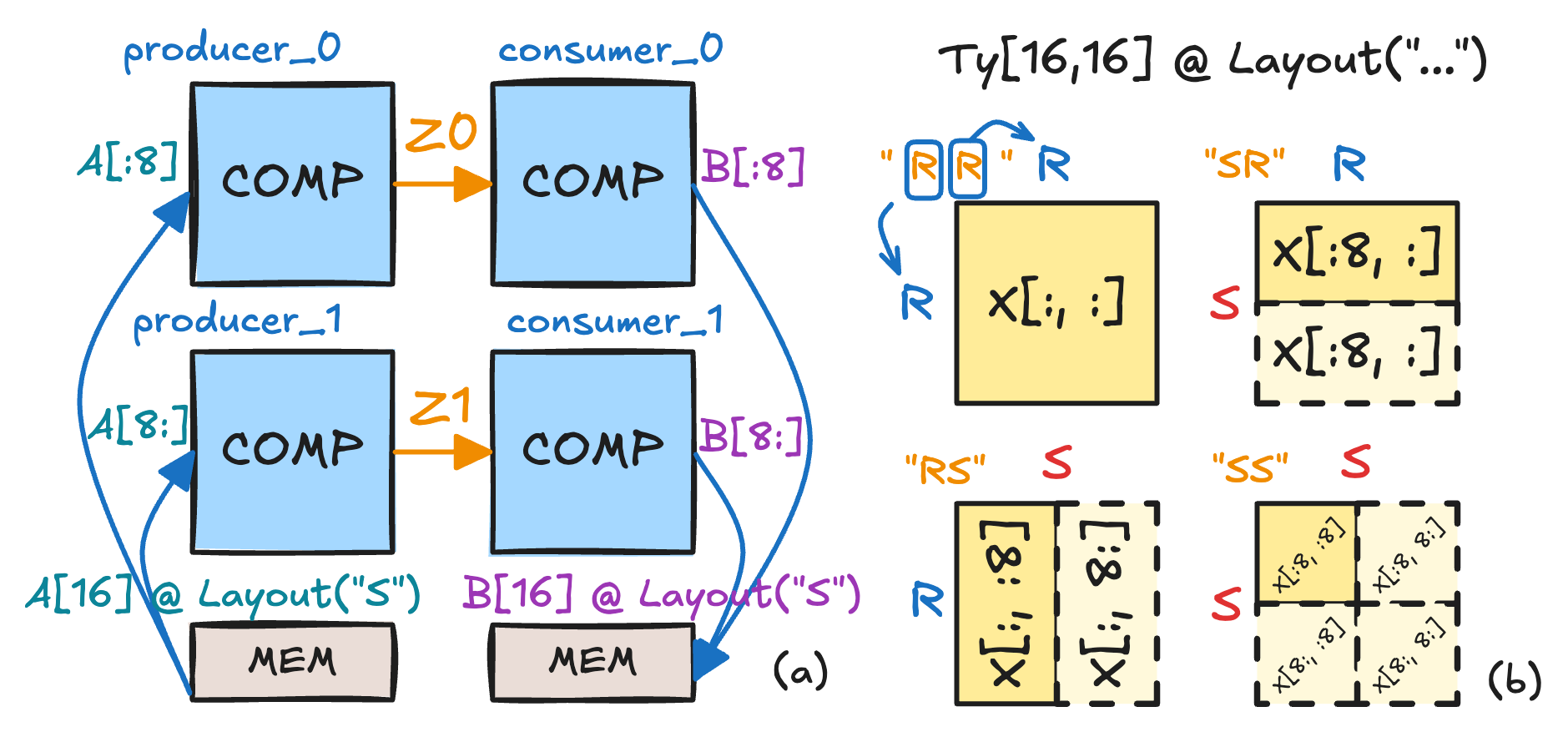}
\caption{(a) Mapping of the program in Fig.~\ref{subfig:dato}. (b) Layout of a 16$\times$16 tensor mapped to a 2$\times$2 mesh.}
\label{fig:memory_type}
% required by acmart, for voiceover to read the paper aloud
\Description{}
\end{figure}

We illustrate a simple producer–consumer pattern written in Allo~\cite{chen2024allo} in Fig.~\ref{subfig:allo}, and the expected mapping is shown in Fig.~\ref{fig:memory_type}a.
We identify several limitations in Allo that hinder productivity and scalability.

\textbf{Limitation \textcircled{1}: Array-based programming introduces semantic indirection.}
In Allo, users must write programs in an array-based style, even though the underlying hardware communicates via FIFOs. Users need to explicitly insert array-to-stream conversions using the \texttt{.relay()} primitive (L23--24), which creates a semantic gap: programmers reason about arrays, while the hardware executes streams. This indirection makes the code less natural to write and more difficult for the compiler to optimize.

\textbf{Limitation \textcircled{2}: Index management relies on tedious manual bookkeeping.}
The Allo code requires explicit partitioning and index slicing (L19--20). Considering different tensors may have different sharding methods for different compute units, such low-level handling is error-prone and becomes unmanageable in larger programs.

\textbf{Limitation \textcircled{3}: Per-PE function definitions do not scale with hardware.}
Allo requires separate functions to be written for each PE (e.g., \texttt{producer\_0}, \texttt{producer\_1} in L6--15). While workable for small examples, this approach scales poorly to accelerators with hundreds of PEs.

To address the limitations above, \Name introduces a task-based programming model, illustrated in Fig.~\ref{subfig:dato}. Users write a top-level function that will be passed into the \Name parser (L8), within which multiple parallel tasks can be defined using \texttt{@task} (L14, L19).
% \zz{is it possible not to mention ``region'' in the paper to simplicity?}
\Name promotes streams to first-class types (L11), allowing programs to explicitly express hardware-level communication and  \textbf{resolve limitation \textcircled{1}}. Sharded inputs are specified via layout refinement types (L14, L19), enabling the compiler to handle partitioning and layout transformations automatically. This effectively eliminates manual index arithmetic and \textbf{addresses limitation \textcircled{2}}. Instead of duplicating code for each processing element, users define a single task and apply virtual mapping (Line 18), letting the compiler replicate and schedule it across the hardware, thus \textbf{overcoming limitation \textcircled{3}}. By aligning abstractions with hardware semantics and automating layout and mapping, \Name eliminates the indirection and bookkeeping required in Allo, streamlining the programming of dataflow accelerators.
% The remaining section will provide details of the stream and layout types.

\subsection{Stream Type}
\label{sub:stream}
% \begin{figure}[!htbp]
% \begin{subfigure}[b]{0.49\linewidth}
% \centering
% \includegraphics[width=\linewidth]{figures/producer-consumer.png}
% \caption{Mapping of the program in Fig.~\ref{subfig:dato}}
% \label{subfig:prod_cons}
% \end{subfigure}
% % \hfill\vline\hfill
% % \begin{subfigure}[b]{0.43\linewidth}
% % \centering
% % \includegraphics[width=\linewidth]{figures/stream.png}
% % \caption{Stream type and its two associated operations}
% % \label{subfig:stream}
% % \end{subfigure}
% % \hfill\vline\hfill
% \centering
% \begin{subfigure}[b]{0.49\linewidth}
% \includegraphics[width=\linewidth]{figures/layout.png}
% \caption{Layout of a 16$\times$16 tensor mapped to a 2$\times$2 mesh}
% \label{subfig:layout}
% \end{subfigure}
% % \hfill
% \caption{Demonstration of the producer-consumer mapping, stream type, and layout type.}
% \label{fig:memory_type}
% % required by acmart, for voiceover to read the paper aloud
% \Description{}
% \end{figure}

We expose a first-class \emph{stream type} in the frontend to model on-chip communication directly, eliminating array-to-stream conversion. Prior work in stream processing typically abstracts streams as unbounded high-level semantic constructs or in multi-producer and multi-consumer settings that are decoupled from hardware capacity constraints~\cite{nick2025parallelstreaming,laddad2025flo,cutler2024streamtype}. Existing hardware-oriented frameworks attempt to formalize streaming but do so via functional or space-time types, relying on run-time handshakes or schedule alignment without static guarantees~\cite{thomas2020fleet,durst2020aetherling}.
In contrast, our stream type is embedded in an imperative Python frontend and faithfully models the semantics of hardware FIFOs and generalizes beyond scalar FIFOs.
% \zz{clarify that this goes beyond scalar FIFOs}

A stream is essentially a first-in–first-out structure that defines a point-to-point connection between two tasks. The type is defined as
$\mathsf{Stream}[T,N,P]$,
where \texttt{T} is the element type, \texttt{N} specifies the logical capacity in elements, and \texttt{P} defines the number of elements bundled per transfer; by default there is no packing.
Two operations \texttt{.put()} and \texttt{.get()} are associated with this type, and can be used to enable communication between tasks, as shown in L16 and L21 in Fig.~\ref{subfig:dato}.
% Making \texttt{Pack} explicit lets programmers align transfers with hardware widths (e.g., vector lanes or DMA beats) and reason about throughput without changing algorithmic code, while \texttt{Dep} captures buffering that enables producer–consumer overlap and latency hiding.

\begin{figure}[!htbp]
\begin{mathpar}
\scalebox{0.9}{
\infer[]{
  \Gamma \vdash s : \Stream{T}{N}
  \qquad
  \Gamma \vdash x : T
  \qquad
  S = \sid(s)
}{
  \Gamma \; ; \; \Delta\uplus \{\,\Free(S)\,\}
  \;\vdash\;
  \mathrm{put}(s,x) : \mathbf{1}
  \; ; \;
  \Delta\uplus \{\,\Used(S)\,\}
}\;\;(\textsc{T-Put})
}
\\
\scalebox{0.9}{
\infer[]{
  \Gamma \vdash s : \Stream{T}{N}
  \qquad
  S = \sid(s)
}{
  \Gamma \; ; \; \Delta\uplus \{\,\Used(S)\,\}
  \;\vdash\;
  \mathrm{get}(s) : \LFut{T}
  \; ; \;
  \Delta\uplus \{\,\Free(S)\,\}
}\;\;(\textsc{T-Get})
}
\\
\scalebox{0.9}{
\infer[]
{ }{
  \Gamma \; ; \; \Delta\uplus\{\, f:\LFut{T}\,\}
  \;\vdash\;
  \mathrm{await}(f) : T
  \; ; \;
  \Delta
}\;\;(\textsc{T-Await})
}
\end{mathpar}
\caption{Typing rules for stream type. $\Gamma$ is the ordinary typing context, whereas $\Delta$ is a multiset representing the linear context that tracks resources that must be used exactly once. $\uplus$ is the multiset union. $N$ is the depth of the stream.}
\label{fig:typing_stream}
\Description{}
\end{figure}

Fig.~\ref{fig:typing_stream} presents the typing rules for streams, which mirror hardware FIFO semantics precisely.
% \zz{not clear what we mean by bounded FIFO; are we considering depth here? without the schedule, we can't really reason about the depth though; need to clarify}
Each element occupies a slot between head (dequeue) and tail (enqueue), and we track these slots with \emph{linear capability tokens} $\Free(S)$ and $\Used(S)$: \texttt{.put()} is well-typed only when a $\Free(S)$ token is available, converting it into $\Used(S)$ (tail advances); symmetrically, \texttt{.get()} is well-typed only in the presence of $\Used(S)$, which it immediately turns back into $\Free(S)$ while returning the payload as a \emph{linear ready-future}, a value available without blocking but required to be consumed exactly \emph{once}.
Freeing the slot at \texttt{.get()} time relieves backpressure as soon as the consumer admits the item, so producers can continue streaming even if the consumer defers use.
We intentionally separate data movement from actual computation and introduce an internal \texttt{.await()} operation to unwrap the future, the actual value to be used.
% \zz{briefly explain the meaning of future and unwrap for reviewers who are not familiar with these PL concepts}
Since the future is ready and linear, the compiler can place the unique \texttt{.await()} at a latency-hiding point without risking duplication or deadlock.
% \zz{we need to clearly state the limitations (e.g., fixed-bound loop only)}
The circulating tokens enforce a simple invariant: each \texttt{.put()} decrements $\Free(S)$ and increments $\Used(S)$ by one, and each \texttt{.get()} does the reverse, so overflow and underflow become untypeable by construction. \texttt{.await()} is FIFO-neutral and merely unwraps an already-dequeued element, preserving both safety and throughput.
% \zz{We need to clarify, in simple terms, the kind of type safety we intend to achieve here.}

\begin{figure}[t]
\begin{subfigure}[b]{0.49\linewidth}
\begin{minted}[linenos,
               fontsize=\scriptsize,
               xleftmargin=1.8em,
               escapeinside=||,
               autogobble]{python}
# |\rxmark| Deadlock
sAB: Stream[T, N]
sBA: Stream[T, N]
@task()
def func0():
 for i in range(N):
  a = sBA.get(); sAB.put(a)
@task()
def func1():
 for i in range(N):
  b = sAB.get(); sBA.put(b)
\end{minted}
\end{subfigure}
\begin{subfigure}[b]{0.49\linewidth}
\begin{minted}[linenos,
               fontsize=\scriptsize,
               xleftmargin=1.8em,
               escapeinside=||,
               autogobble]{python}
# |\rxmark| Inconsistent put/get
s: Stream[T, N]; a = 0
@task()
def func0():
 for i in range(N):
  s.put(a)
  s.put(a)
@task()
def func1():
 for i in range(N):
  b = s.get()
\end{minted}
\end{subfigure}
% \hfill\vline\hfill
% \begin{subfigure}[b]{0.49\linewidth}
% \begin{minted}[linenos,
%                fontsize=\scriptsize,
%                xleftmargin=1.8em,
%                highlightlines={11,14,18},
%                escapeinside=||]{python}
% S: Stream[T, N]
% @task()
% def sw_pipeline():
%  cur = S.get()
%  for i in range(1, N):
%   next = S.get()
%   y = compute(next)
%   cur = next
%  y = compute(cur)
% \end{minted}
% \end{subfigure}
\caption{Example dataflow programs that fail type check.}
\label{fig:type_check}
% required by acmart, for voiceover to read the paper aloud
\Description{}
\end{figure}

% A correct pipelined loop (accepted)
% with fifo_scope(s_in, capacity=4) as (P, C):
%     f_cur = C.get()                 # needs Used(s_in); returns Free(s_in) + LFut[T]
%     while C.can_get():              # refines: Used(s_in) ≥ 1 at loop head
%         f_next = C.get()            # Used→Free immediately (frees a slot early)
%         do_control_work()           # doesn’t need the data yet
%         x = await f_cur             # consume linear future; no FIFO effect
%         y = compute(x)
%         sink.put(y)                 # needs Free(sink); normal
%         f_cur = f_next
%     # epilogue
%     y = compute(await f_cur)
%     sink.put(y)

% Per iteration, the net token effect is zero (one get gives back one Free, one await is FIFO-neutral), so the loop is Φ-preserving and type-checks.

To type check programs using streams, \Name initializes each $\Stream{T}{N}$ with a multiset of $N$ tokens $\Free(S)$ and zero $\Used(S)$. At every program point, we maintain an abstract state mapping streams to their current token multisets. Type checking is performed via forward abstract interpretation~\cite{kildall1973datflow} over the control flow graph (CFG). A program is well-typed if, upon exiting each region, all futures are consumed and no tokens are leaked.
Fig.~\ref{fig:type_check} shows several untypeable programs that are rejected by our type system.
We also note that this type system is an untimed model, so it cannot provide timing information (e.g., minimal FIFO depths) that depends on concrete hardware scheduling.

By making streams a first-class type in \Name, users can naturally express higher-level patterns such as arrays of streams or streams of tensors. This composability scales to multi-stage pipelines and fork-join topologies, enabling clear and analyzable communication while retaining a direct mapping to hardware. Compile-time checking not only guarantees safety but also provides early feedback to help performance engineers debug and optimize their designs. We provide more concrete examples in \S~\ref{sub:multikernel_exp} to illustrate this compositional expressiveness in practice.

\subsection{Layout Type}
\label{subsec:layout}
As deep learning scales across distributed, heterogeneous systems, treating individual PEs or devices as first-class programming units quickly becomes unmanageable. The single-program multiple-data (SPMD) paradigm therefore serves as the prevailing foundation, expressing one program over partitioned data while the system orchestrates placement and communication. Building on SPMD, recent work on annotated sharding~\cite{zheng2022alpa,xu2021gspmd,chen2024slapo,xie2022p2,sami2025partir} provides (semi-)automated mechanisms to partition tensors and schedule computation for large models.
A similar challenge arises on dataflow accelerators with hundreds of PEs. As discussed in \S~\ref{sub:motivation}, manually computing per-tile index arithmetic and ensuring kernel-level consistency is brittle and time-consuming. We take the next step by promoting data layout to a first-class static construct: the layout across PEs is encoded directly into the type system as a \emph{refinement} of the base tensor type, denoted as
\texttt{T[Shape] @ Layout("...")}
in Python (L14, L19 in Fig.~\ref{subfig:dato}).
% \zz{Use L14, L19 also in other places}
% Users may omit this refinement if they do not require specific sharded layouts.

\begin{figure}[t]
\begin{mathpar}
% Labels, layouts, shapes, types, effects
\ell ::= \mathbf{S} \mid \mathbf{R}
\quad
L ::= \langle \ell_1,\dots,\ell_d\rangle \in \{\mathbf{S},\mathbf{R}\}^d
\quad
\tau ::= D[\vec n] @ L
\\
\infer[Well-Formedness]{|L|=\mathrm{rank}(\vec n)}{\Gamma \vdash D[\vec n]@L}
\quad
% \text{Per-axis join:}\quad
\ell_1 \sqcup \ell_2 =
\begin{cases}
\mathbf{R} & \text{if } \ell_1=\mathbf{R}\ \wedge\ \ell_2=\mathbf{R}\\
\mathbf{S} & \text{otherwise}
\end{cases}
\\
(L_1 \sqcup L_2)[i] = L_1[i] \sqcup L_2[i]
\\

% ----- Elementwise -----
\infer[]{
\Gamma \vdash x : D[\vec n]@L_x\ !\ \Pi_x \quad
      \Gamma \vdash y : D[\vec n]@L_y\ !\ \Pi_y}
     {\Gamma \vdash x \odot y : D[\vec n]@(L_x \sqcup L_y)\ !\ (\Pi_x \cup \Pi_y)}
\;\;(\textsc{Eltwise})
\\
% ----- Reduction along axis i by }\oplus\text{ -----
\scalebox{0.92}{
\infer[]{
\Gamma \vdash e : D[\vec n]@L\ !\ \Pi \quad L[i]=\mathbf{S}}
     {\Gamma \vdash \mathrm{reduce}_{\oplus,i}(e) :
        D[\vec n\setminus n_i]@ (L\setminus L[i])\ !\ (\Pi \cup \{\oplus\})}
\;\;(\textsc{Reduce-S})
}
\\
\infer[]{
\Gamma \vdash e : D[\vec n]@L\ !\ \Pi \quad L[i]=\mathbf{R}}
     {\Gamma \vdash \mathrm{reduce}_{\oplus,i}(e) :
        D[\vec n\setminus n_i]@ (L\setminus L[i])\ !\ \Pi}
\;\;(\textsc{Reduce-R})
\\
% ----- Matrix multiply A:(m×k), B:(k×n) -----
\scalebox{0.85}{
\infer[Matmul]{
\Gamma \vdash A : D[m,k]@(\ell_a^1,\ell_a^2)\ !\ \Pi_A \quad
      \Gamma \vdash B : D[k,n]@(\ell_b^1,\ell_b^2)\ !\ \Pi_B \quad
      \ell_a^2 = \ell_b^1}
     {\Gamma \vdash \mathrm{matmul}(A,B) :
        D[m,n]@(\ell_a^1,\ell_b^2)\ !\ (\Pi_A \cup \Pi_B \cup (\{+\}\ \text{iff }\ell_a^2=\mathbf{S}))}
}

% ----- Resolve pending partials -----
\infer[]{
\Gamma \vdash x : \tau\ !\ (\Pi \cup \{\oplus\})}
     {\Gamma \vdash \mathrm{allreduce}_{\oplus}(x) : \tau\ !\ \Pi}
\;\;(\textsc{AllReduce})
\end{mathpar}
\caption{Layout type definition and minimal typing rules. $D$ is the base element domain. $\Gamma$ is the typing context. $\tau\ !\ \Pi$ denotes pending-collective effects $\Pi$ to a type $\tau$.}
\label{fig:typing_layout}
\Description{}
\end{figure}

% \zz{Add a comment in Fig 6 L5 -- S1 means sharding dim 1}
Fig.\ref{fig:typing_layout} defines the layout type and its typing rules.
Each tensor axis carries a layout label, either \textbf{R} (replicated) or \textbf{S} (sharded).
An example layout for a 2D tensor is shown in Fig.\ref{fig:memory_type}b.
Reducing along the axis $i$ eliminates that axis from the shape in both cases, but the information available on a single shard differs. If 
$L[i]=\mathbf{R}$ (\textsc{Reduce-R}), every shard already holds the full slice of axis $i$; the reduction is therefore purely local and final, so the effect set remains unmodified.
If $L[i]=\mathbf{S}$ (\textsc{Reduce-S}), each shard only sees its partition of axis $i$; a local reduction produces a partial result that must still be combined across shards. We record this obligation by adding the operator $\oplus$ (e.g., $+$, $\max$) to the effect set $\Pi$.
% Thus, \textsc{Reduce-R} and \textsc{Reduce-S} perform the same structural change (drop axis $i$ and its label), but only \textsc{Reduce-S} introduces a pending collective.

The pending effect is discharged by a collective that does not change shape or layout. The rule for $\mathrm{allreduce}_\oplus$ simply removes $\oplus$ from $\Pi$, meaning the per-shard partials have been globally combined and the value is now materialized with respect to $\oplus$. This design cleanly separates algorithmic intent (the tensor reduction over an axis) from distributional synchronization (the cross-shard combine), and it preserves composability: partial results can flow through subsequent operations, accumulating or being preserved in $\Pi$, until a consumer requires a materialized value.
% or the compiler’s cost model chooses to insert a collective.
% In practice, the \Name compiler fuses local computation before paying for communication, while keeping the contract explicit in the types: reducing an $S$ axis creates debt ($!\{\oplus\}$), and an $\mathrm{allreduce}_\oplus$ clears it without further reshaping.

By default, the layout type is $\mathbf{R}^d$, indicating that all dimensions are replicated. This means users can write vanilla tensor code without additional layout annotations unless needed. Attaching layout as a refinement type enables static checking of layout compatibility, prevents unsound or implicit re-sharding, and allows \Name to automatically insert collectives and provide precise diagnostics when inconsistencies arise. This design preserves the familiar syntax of tensor programs while enriching the type system with layouts, offering users fast and actionable feedback with minimal overhead.
\section{Virtual Mapping}
\label{sec:mapping}
In this section, we introduce the mapping primitives for scaling up the design and discuss techniques for automating the virtual-to-physical mapping process.

\begin{figure*}[t]
\begin{subfigure}[b]{0.29\linewidth}
\begin{minted}[linenos,
               fontsize=\scriptsize,
               xleftmargin=1.8em,
               escapeinside=||,
               autogobble]{python}
import dato
from dato.ir.types import Layout

Pk, Pm, Pn = 2, 2, 2
LyA = Layout("S1S2")
LyB = Layout("S2S0")
LyC = Layout("S1S0")

def top():
  @dato.task(mapping=[Pk, Pm, Pn])
  def gemm(A: Ty[M, K] @ LyA,
           B: Ty[K, N] @ LyB,
           C: Ty[M, N] @ LyC):
    part_C = dato.matmul(A, B)
    C[:, :] = \
      dato.allreduce(part_C, op="+")
\end{minted}
\caption{A tiled GEMM kernel in \Name}
\label{subfig:tiled_gemm}
\end{subfigure}
\begin{subfigure}[b]{0.4\linewidth}
\centering
\includegraphics[width=\linewidth]{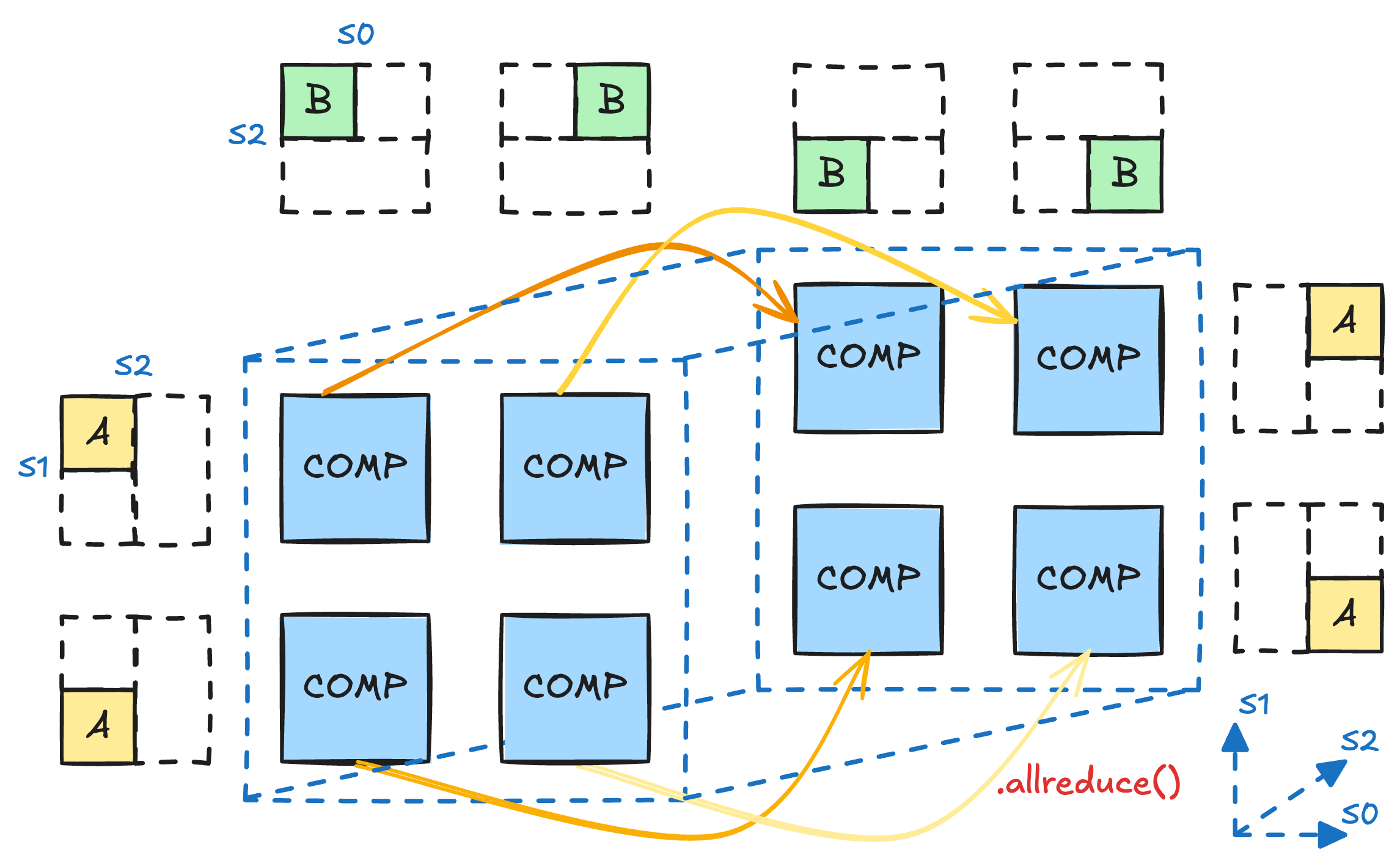}
\caption{Virtual mapping of the GEMM kernel}
\label{subfig:gemm_mapping}
\end{subfigure}
\begin{subfigure}[b]{0.3\linewidth}
\centering
\includegraphics[width=\linewidth]{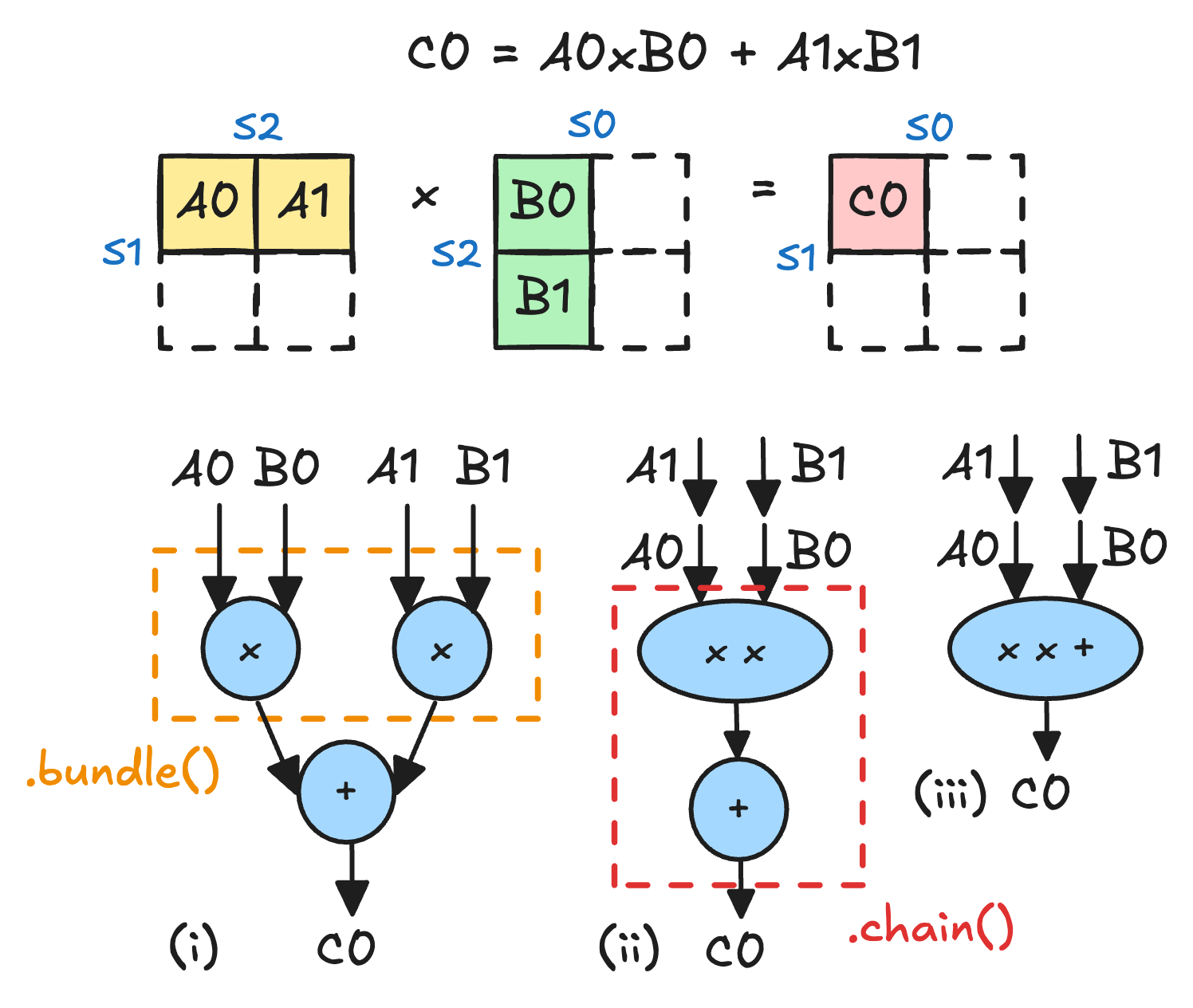}
\caption{\texttt{.bundle()} and \texttt{.chain()} primitives}
\label{subfig:bundle_chain}
\end{subfigure}
% \hfill
\caption{Virtual-to-physical mapping of a tiled GEMM kernel.}
\label{fig:virtual_mapping}
% required by acmart, for voiceover to read the paper aloud
\Description{}
\end{figure*}

\subsection{Virtual Mapping Graph}
When a single tiling dimension exceeds on-chip capacity, \Name naturally generalizes to multi-dimensional tiling and mapping. As shown in L5--7 of Fig.~\ref{subfig:tiled_gemm}, layout annotations specify how each tensor dimension is sharded and which device axis it maps to, indicating the eventual placement of that dimension in the spatial fabric (e.g., \texttt{S1S2} means dim 0 of A is sharded and mapped to device axis 1, while dim 1 is sharded and mapped to device axis 2). Given the mapping directive (L10), \Name instantiates a 3D lattice of task instances---one instance per point in the virtual grid.
% \zz{one instance per point?}
The resulting virtual mapping is illustrated in Fig.~\ref{subfig:gemm_mapping}.

To capture the actual computation and enable collapsing of virtual computation onto physical PEs, we construct a virtual mapping graph (VMG) by wiring task instances according to the program's stream operations and layout annotations. Each \texttt{task} body becomes a node in the VMG, representing a PE-level kernel (e.g., a local \texttt{matmul}). Directed edges are introduced by communication primitives: every \texttt{.put()}/\texttt{.get()} on a \texttt{Stream} establishes a producer-to-consumer edge. Collective operators such as \texttt{.allreduce()} generate reduction edges, connecting all task instances that participate in the collective as dictated by the layout. For example, when the reduction dimension is sharded along a virtual axis, \texttt{.allreduce()} aggregates the partial results (e.g., \texttt{part\_C}) into the final output (e.g., \texttt{C}), as shown in L14--16 of Fig.~\ref{subfig:tiled_gemm}.
The resulting VMG is device-agnostic and is visualized in Fig.~\ref{subfig:bundle_chain}(i), which depicts the subgraph rooted at $C_0$. Similar subgraphs rooted at $C_1$ through $C_3$ share the same structure.

\subsection{Mapping Primitives}
To map each node in VMG to the actual physical PEs on hardware, we introduce two scheduling primitives \texttt{.bundle()} and \texttt{.chain()} that reshape graph topology while preserving program semantics.
These primitives are exposed to users for manual control of the virtual-to-physical mapping, but they can also be applied automatically by \Name (\S~\ref{sub:automation}).

The \texttt{.bundle()} primitive merges a set of isomorphic nodes (i.e., those with identical computation and I/O patterns) into a single multi-shot node. The resulting node is scheduled to execute multiple times, once per original task, handling each input or output in turn. Conceptually, \texttt{.bundle()} transforms explicit fan-in/fan-out into a time-multiplexed join or split. This reduces the number of physical nodes and external ports, eliminates intermediate buffers, and is especially effective on hardware with limited routing or port availability but sufficient compute capacity to host multiple invocations on the same tile.

The \texttt{.chain()} primitive fuses two connected nodes---typically a producer–consumer pair---into a single sequential node. The fused node preserves all external edges and orders the computations to respect their data dependency. This often removes the intermediate stream and its associated buffering. By collapsing a path into a single location, \texttt{.chain()} improves locality and reduces routing pressure. The fused node must still obey the platform's port constraints (e.g., a maximum of two global inputs and outputs, or shared ports for compatible types).

An example is shown in Fig.~\ref{subfig:bundle_chain}, illustrating the partial-sum pattern
$C_0 = A_0B_0 + A_1B_1$.
Starting from the initial VMG in (i), applying \texttt{.bundle()} to the parallel multiply nodes $A_0B_0$ and $A_1B_1$ merges their outputs into a single join node that accumulates the results into $C_0$, as shown in (ii). Further applying \texttt{.chain()} with the adder yields a single fused node that computes $A_0B_0$ and $A_1B_1$ sequentially and performs the accumulation internally, as shown in (iii). Externally, this graph exposes only two global inputs (the $A$ and $B$ streams reused across shots) and one output for $C_0$. This variant minimizes routing and buffering overhead at the cost of reduced inter-node parallelism.

\subsection{Automated Mapping Flow}
\label{sub:automation}
\begin{algorithm}[!htbp]
\caption{Search for optimal primitive sequences}
\label{alg:bundle-chain-search}
\small
\KwIn{Initial VMG $G$; Available physical PE count $C$}
\KwOut{Optimized module $m_\mathsf{opt}$}

\SetKwFunction{Span}{Span}
\SetKwFunction{IsValidBundleOrChain}{is\_valid\_bundle\_or\_chain}
\SetKwFunction{IsValidBundle}{is\_valid\_bundle}
\SetKwFunction{IsValidChain}{is\_valid\_chain}
\SetKwFunction{ApplyBundle}{apply\_bundle}
\SetKwFunction{ApplyChain}{apply\_chain}
\SetKwFunction{ApplyPrimitive}{apply\_primitive}
\SetKwFunction{Build}{build}
\SetKwFunction{Optimize}{apply\_optimization\_passes\_on}
\SetKwFunction{FindBest}{find\_optimal\_among}

\BlankLine
\textbf{global} $\mathcal{M} \gets \varnothing$ \tcp*{mapping\_candidates}

\SetKwProg{Fn}{Function}{}{}
\Fn{\Span{state}}{
  \If{$state.\mathrm{v\_node\_number} \le C$}{
    \Optimize{$state.\mathrm{vmg}$}\;
      \If{$\mathsf{module} \leftarrow$ \Build{$state.\mathrm{vmg}$} \text{succeeds}}{
        $\mathcal{M} \gets \mathcal{M} \cup \{\mathsf{module}\}$\;
      }
      \If(\quad\tcp*[h]{early\_stop}){$|\mathcal{M}|>\mathsf{threshold}$}{
        \Return
      }
  }
  \ForEach(\tcp*[h]{priority-based expansion}){$(u,v) \in state.\mathrm{v\_node\_pairs}$}{
    \If{\IsValidBundleOrChain{$u,v$}}{
      $t \gets state.$\ApplyPrimitive{$u,v$}\; \Span{$t$}\;
    }
  }
}

$root \gets \mathrm{SearchTree}(G)$; \Span{$root$}\;
$m_{\mathsf{opt}} \gets $\FindBest{$\mathcal{M}$}\;
\Return $m_{\mathsf{opt}}$\;
\end{algorithm}

% $\mathcal{V} \gets [\ ]$ \tcp*{valid\_results}
% \ForEach{$m \in \mathcal{M}$}{
%   \Optimize{$m$}\;
%   \If{$\exists\,\mathsf{module} \leftarrow$ \Build{$m$} \text{ succeeds}}{
%     append $\mathsf{module}$ to $\mathcal{V}$\;
%   }
%   % \tcp*{ignore failures (e.g., memory/ports/mapping)}
% }
% $\mathsf{opt} \gets$ \FindBest{$\mathcal{V}$}\; \Return $\mathsf{opt}$\;
When the VMG is large, manually collapsing the graph onto physical PEs becomes tedious and error-prone. To address this, we propose a search algorithm that generates a sequence of mapping primitives along with the corresponding physically mapped program. As shown in Alg.~\ref{alg:bundle-chain-search}, the algorithm maps a VMG $G$ onto a compute fabric with a resource budget of $C$ by systematically ``collocating'' nodes until the number of virtual nodes falls within the available budget. Mapping primitives are applied only when lightweight, architecture-aware legality conditions are satisfied. These conditions are governed by three main criteria:
\textbf{(i) Resource constraints}, such as top-level I/O port availability, per-tile ingress and egress limits on NPUs, and BRAM access limits on FPGAs;
\textbf{(ii) Interface compatibility}, requiring that the nodes have matching element types, shapes, and producer–consumer relationships to ensure external behavioral equivalence;
\textbf{(iii) Dependency soundness}, allowing only nodes that are either independent or directly connected by a producer–consumer edge, in order to avoid cycles that might arise from collapsing across transitive paths.
For I/O accounting, only global ports are constrained. Local communication within collocated nodes is treated as internal and does not count against global port budgets---for instance, tile-local links on NPUs or on-chip wiring in FPGAs. Under these rules, each collocation step reduces the number of concurrently active nodes while adhering to platform constraints on ports, routing, and memory, making the algorithm portable across different targets.

At a high level, the procedure performs a branch-and-reduce search over a tree rooted at the initial VMG. Each node in this tree represents a mapping graph, or \emph{state}, resulting from a sequence of primitive applications. For instance, the three graphs shown in Fig.~\ref{subfig:bundle_chain} illustrate distinct states within this search tree. Each edge corresponds to a single legal primitive application that strictly reduces the number of virtual nodes, ensuring monotonic descent. From any intermediate state, the algorithm enumerates candidate node pairs and filters them using the legality conditions described earlier. Admissible pairs yield child states, which are explored recursively. States with at most $C$ virtual nodes are recorded as feasible mappings, while those that still exceed the resource budget and allow no further valid transformation are pruned.
To guide the exploration, we incorporate a priority function that favors transformations likely to improve hardware compatibility---for example, by reducing global I/O usage or relieving port pressure. This heuristic ordering improves convergence without sacrificing correctness. Once a feasible mapping is found, the candidate undergoes target-specific optimization and code generation (\S~\ref{sec:opt}). Mappings are evaluated until a desired number of valid implementations is reached. Finally, the mapping with the highest measured performance is selected as the output.

\section{Further Optimizations}
\label{sec:opt}
In this section, we discuss further optimizations that are necessary for dataflow accelerators.

\subsection{Kernel Injection}
Many NPUs ship hand-tuned C++ vector kernels for their VLIW cores~\cite{hunhoff2025iron,mliraie}, while FPGAs offer pre-verified C++ HLS IP blocks for reuse~\cite{vitis_ai,vitis_hls_library}. To exploit these library functions, we define a kernel interface with an explicit contract: (i) admissible shapes and vector widths, (ii) required I/O layouts (e.g., tiled or packed), and (iii) cost/resource hints such as latency and initiation interval---augmented with device specifications for each target. When a computation graph satisfies a contract, \Name's lowering pass selects an appropriate variant and synthesizes a wrapper that tiles task loops to the kernel's micro-tiling, inserts prologue/epilogue layout adaptors, and stitches streams and DMAs to overlap packing/unpacking with compute. On NPUs, the wrapper enforces operand layouts and vector widths to fully leverage VLIW execution; on FPGAs, it emits HLS-friendly loops with explicit pipelining around the kernel call.

% \fixme{These contracts are recorded in \Name to enable legality checks and pattern-based rewrites during lowering.} \zz{rephrase -- do we have a device spec that defines the target architecture? how do we distinguish NPUv1 from v2?}
% Concretely, the pass emits an MLIR \texttt{func.call} to the external kernel accompanied by: (a) vectorized pack/unpack loops or layout-aware \texttt{dma2d} ops, (b) double- (or triple-) buffered local memories to hide adaptor latency, and (c) async tokens to pipeline {pack; call; unpack} across tiles.
% The net effect is to reduce high-level ops to performant library calls without forfeiting global pipeline structure.

\subsection{Fine-Grained Layout Optimization}
We further introduce a fine-grained layout optimization that operates within a single PE, given that high-performance kernels often impose strict input/output layout requirements. Following CuTe~\cite{cute} and Graphene~\cite{hagedorn2023graphene}, we represent tensor layouts explicitly by recording the tuple $(\mathsf{offsets}, \mathsf{sizes}, \mathsf{strides})$. Naive lowering around injected kernels tends to leave behind chains of small reformatting steps---tile, pack, transpose, re-pack---that inflate bandwidth and stress local memories. To address this, we develop two complementary optimizations: \emph{normalization \& collapse} and \emph{DMA-aware hoisting}.

\textbf{Normalization \& collapse.} We model layout transformations as elements of a layout algebra~\cite{zhou2025linearlayout,cute,hagedorn2023graphene}: compositions of affine maps over integer domains subject to tiling divisibility constraints. \Name lifts each sequence of MLIR \texttt{transform\_layout} operations into this algebra, canonicalizes the composition (e.g., merging adjacent blocks, cancelling inverse pairs), and then \emph{collapses} it to a minimal normal form. Concretely, we maintain the layout tuple in SSA, compose them symbolically, and discharge integrality side conditions with lightweight arithmetic checks. The result eliminates redundant materializations and fuses compatible steps into a single pack/unpack loop or a single DMA descriptor. In contrast to local peepholes~\cite{chen2018tvm,xla}, algebraic normalization exposes global cancellation opportunities across loop boundaries and between producer–consumer tasks.

\textbf{DMA-aware hoisting.} Many layout operations such as packing, interleaving, striding, and fixed-pattern permutations, can be executed \emph{in flight} by NPU DMA engines. Our pass \emph{hoists} eligible layout work across stream boundaries and folds it into DMA descriptors: a pack becomes a strided gather and a transpose becomes a 2D stride swap. Hoisting is guarded by legality checks on stride ranges, burst sizes, and bank mappings, improving effective bandwidth and freeing local SRAM. On FPGA backends, the same transformation lowers to HLS-friendly burst-aligned loops. Finally, the scheduler places these DMAs in producer- and consumer-initiated slots and relies on multi-buffering to keep compute saturated while the engine performs the transformation.

\section{Implementation}
\label{sec:impl}
\Name is built atop Allo~\cite{chen2024allo}, comprising 4K lines of Python and 1K lines of C++ for the frontend and common optimization passes. An additional 3K lines of code are dedicated to high-performance NPU code generation. The compiler leverages the MLIR infrastructure to target multiple hardware platforms, emitting MLIR-AIE for NPUs and C++ HLS for FPGAs.

Apart from the optimizations discussed in \S~\ref{sec:opt}, we implement a DMA scheduling pass to address I/O bottlenecks on dataflow accelerators, where each PE typically exposes only a few I/O ports (e.g., 2 in/2 out). This becomes a challenge when virtual mapping inflates the number of live arguments, risking FIFO contention and deadlocks.
To manage this, we introduce a token-based scheduling strategy grounded in a coarse liveness analysis. Rather than tracking exact lifetimes across tiles, we conservatively approximate each argument's live interval using the span of its originating function. Based on this abstraction, we group transfers into epochs, each corresponding to a disjoint set of arguments. \texttt{.chain()} serialize function instances across epochs, while \texttt{.bundle()} assigns distinct tokens to parallel replicas. These tokens guide the construction of DMA transfers, which are then scheduled per tile under fixed port budgets. Ports are acquired at first use and released at last use, ensuring arguments that share ports have non-overlapping lifetimes.
We further enhance port utilization through three key optimizations: \textbf{(1) multicast merging}, which consolidates duplicate transfers of the same tile to multiple destinations; \textbf{(2) spatial coalescing}, which fuses adjacent transfers with matching shapes and strides; and \textbf{(3) port-aware splitting}, which divides large coalesced DMAs to fit within the available ports per epoch. The resulting schedule maintains correctness while achieving high throughput.
% More advanced analyses, such as precise aliasing and cross-tile deadlock checks, can be layered on without altering the scheduling interface.
\section{Experiments}
\label{sec:exp}
In this section, we first discuss our experimental setting and evaluate our compilation flow.

\begin{figure*}[t]
\includegraphics[width=\linewidth]{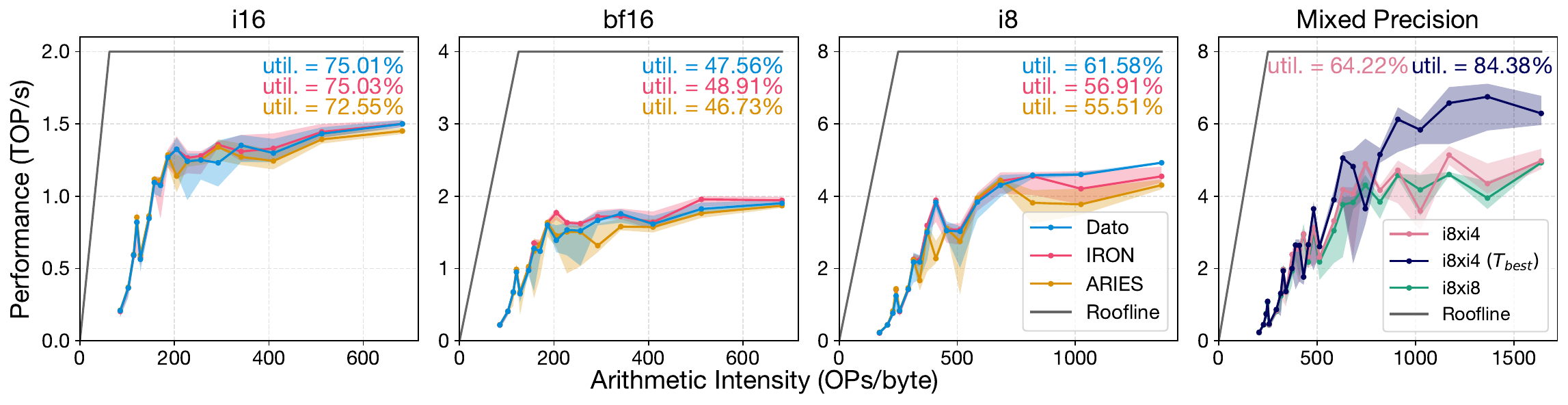}
\caption{Performance results on the GEMM kernel with different data types. Shaded regions denote the execution variance.}
\label{fig:gemm_result}
% required by acmart, for voiceover to read the paper aloud
\Description{}
\end{figure*}

\subsection{Experimental Settings}
% \zz{provide the reason why we pick these two kinds of devices}
We evaluate \Name on both NPUs and FPGAs, representing commercial coarse-grained and fine-grained dataflow accelerators, respectively. NPUs feature fixed-function PE arrays optimized for VLIW execution, while FPGAs offer LUT-based reconfigurability, enabling custom dataflow structures. For NPU evaluation, we compare against IRON~\cite{hunhoff2025iron}, the official AMD flow built on MLIR-AIE~\cite{mliraie}, and ARIES~\cite{zhuang2025aries}, a state-of-the-art AIE compilation framework.
Experiments are conducted on the AMD Ryzen AI NPU (XDNA1) with 20 compute tiles ($4 \times 5$). For single-kernel benchmarks, we report on-device kernel execution time, excluding one-time host-device transfer latency, to better isolate hardware utilization---defined as the ratio of actual compute throughput to the theoretical peak.
% Experiments are conducted on two generations of AMD Ryzen AI NPUs: XDNA1 with 20 compute tiles ($4\times 5$) and XDNA2 with 32 compute tiles ($4\times 8$). In the single-kernel GEMM experiments, we report on-device kernel execution time, excluding one-time host-device transfer latency to isolate hardware utilization (defined as the ratio of the actual computational work to the hardware theoretical peak capacity).
% \zz{we still need to properly define utilization; stick to either compute or hardware utilization; is the host-device transfer one-time effort? otherwise, the reviewers would surely challenge this setting}
For multi-kernel workloads, we measure end-to-end latency, including data movement overheads. Each benchmark includes 20 warm-up iterations followed by 400 timed runs, reporting the arithmetic mean.
For FPGA evaluation, we target the AMD Alveo U280 board and compare against Allo~\cite{chen2024allo}. Designs are synthesized using Vitis 2023.2~\cite{vitis_hls_2023} with latency taken from the HLS reports.
% The designs are further pushed to backend synthesis, and the frequency and resource utilization results are after placement and routing.

% \zz{we are missing evals on the new programming constructs; currently, it's not clear at all how the new types are used and whether they are useful}

\subsection{Overall Performance on NPU}
First, we measure the overall performance of \Name on both single-kernel and multi-kernel designs.

\subsubsection{Single-Kernel Design}
The single-kernel GEMM benchmark sweeps matrix sizes $M,N,K\in\{256,512,1024,2048\}$ (64 total configurations). We map the kernel to a $4\times 4$ active-tile region on NPU and evaluate three precisions: \texttt{i16}, \texttt{bf16}, and \texttt{i8}. Tiling is fixed per setting: for \texttt{i16}/\texttt{bf16} we use $m$=$n$=$k$=$64$; for \texttt{i8} we use $m$=$k$=$64$ and $n$=$128$. Fig.~\ref{fig:gemm_result} shows the roofline models. Across precisions, throughput increases with arithmetic intensity and then saturates at the compute ceiling. On \texttt{i16}, \Name matches the best vendor flow at 75.01\% of the peak performance. On \texttt{bf16}, \Name attains 47.56\%, slightly below IRON yet above ARIES. The largest margin appears on \texttt{i8}, where \Name reaches 61.58\%, outperforming both IRON and ARIES; the latter lacks DMA scheduling, leading to a pronounced gap at high intensity. These trends hold over the entire intensity range, with \Name remaining on or near the top curve from the bandwidth-bound regime through the plateau.

The results indicate that \Name's optimizations can match or exceed the performance of hand-crafted templates in IRON. We attribute the advantage to three factors: (i) virtual mapping that scales cleanly with tiling and balances thousands of tasks, a prerequisite for large GEMMs; (ii) communication-aware DMA scheduling with liveness-guided port reuse, which sustains data supply under the NPU's 2-in/2-out constraints; and (iii) vectorized micro-kernels coupled with layout normalization/collapse and DMA-aware hoisting, aligning loop tiles with the kernel's vector width and minimizing pack/transpose overhead. Ablation studies in \S~\ref{sub:ablation} quantify the contribution of each component.

Moreover, \Name enables mixed-precision capabilities that are not supported by other frameworks. As shown in Fig.~\ref{fig:gemm_result}, using \texttt{i4} and \texttt{i8} tensors allows larger tiling configurations, ultimately achieving up to 84\% hardware utilization by fully leveraging the lower bitwidth.

\subsubsection{Multi-Kernel Design}
\label{sub:multikernel_exp}
We evaluate two multi-kernel designs representative of Transformer workloads. The first is multi-head attention (MHA)~\cite{vaswani2017transformer}, realized as two GEMMs with a numerically stable softmax between them; the second is the feed-forward network (FFN), where we implement the up- and gate-projections used in LLaMA-style SwiGLU blocks~\cite{dubey2024llama3}.
We further implement a FlashAttention~\cite{dao2022flashattention} kernel in \Name, which, to the best of our knowledge, is the first demonstration of FlashAttention on AMD NPUs using a high-level programming framework.
We set sequence length $L\in[128, 2048]$ with 12 heads and head dimension 64; the FFN uses hidden sizes 768 and 3072.
ARIES does not provide a usable multi-kernel implementation, so we compare \Name only against IRON. All experiments use \texttt{bf16} on NPU with $4\times 4$ tiles active. Because IRON exposes fixed, per-kernel templates, kernels must be invoked sequentially, forcing temporal execution between stages.

\begin{figure}[t]
\includegraphics[width=\linewidth]{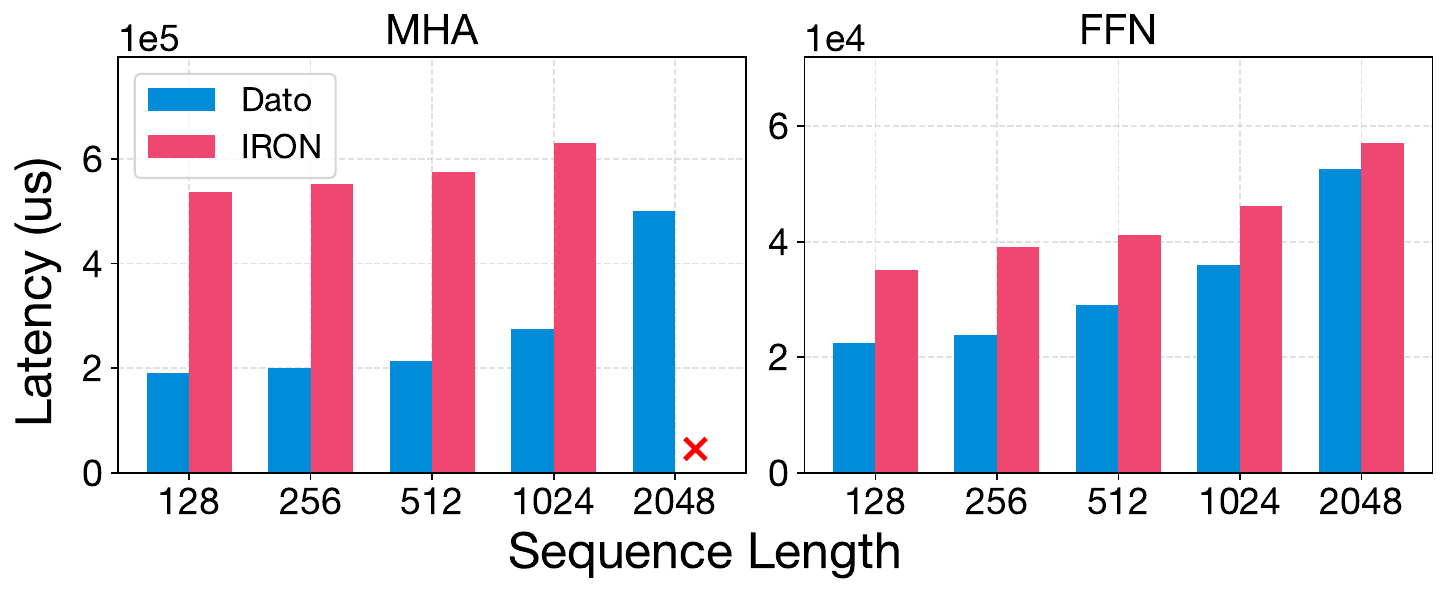}
\caption{Performance of MHA and FFN on NPU (\texttt{bf16}).}
\label{fig:mha_ffn_result}
% required by acmart, for voiceover to read the paper aloud
\Description{}
\end{figure}
\begin{figure*}[t]
\begin{subfigure}[b]{0.64\linewidth}
\begin{minipage}[b]{0.46\linewidth}
\begin{minted}[linenos,
         fontsize=\scriptsize,
         xleftmargin=1.8em,
         escapeinside=||,
         autogobble]{python}
import dato
from dato.ir.types import \
  bfloat16, Stream, Layout

Ty = bfloat16
P0, P1 = N // Tq, N // Tkv
LyQ = Layout("S1R")
LyK = Layout("RS0")
LyV = Layout("S0R")

def top():
  q: Stream[Ty[Tq,d]][P0,P1]
  s: Stream[Ty[Tq,Tkv]][P0,P1]
  w: Stream[Ty[Tq,Tkv]][P0,P1]
  exp_sum: Stream[Ty[Tq]][P0]
  exp_scale: Stream[Ty[Tq]][P0]
  o: Stream[Ty[Tq,d]][P0,P1]
  
  @dato.task(mapping=[P0,1])
  def send_q(Q: Ty[N,d] @ LyQ):
    ti, _ = dato.get_tid()
    for i in range(P1):
      q[ti,i].put(Q)
  
  @dato.task(mapping=[P0,P1])
  def gemm0(K: Ty[d,N] @ LyK):
    ti, tj = dato.get_tid()
    s[ti,tj].put(dato.matmul(
      q[ti,tj].get(), K))
\end{minted}
\end{minipage}\hfill
\begin{minipage}[b]{0.54\linewidth}
\begin{minted}[linenos,
         fontsize=\scriptsize,
         firstnumber=last,
         xleftmargin=1.8em,
         escapeinside=||,
         autogobble]{python}

  @dato.task(mapping=[P0,1])
  def softmax():
    ti, _ = dato.get_tid()
    m: Ty[Tkv]
    sum: Ty[Tkv]
    init_softmax(m,sum)
    for i in range(P1):
      weight: Ty[Tq,Tkv]
      scale: Ty[Tkv]
      online_softmax(
        s[ti,i].get(), m, sum, weight, scale)
      exp_scale[ti].put(scale)
      w[ti,i].put(weight)
      exp_sum[ti].put(sum)
      
  @dato.task(mapping=[P0,P1])
  def gemm1(V: Ty[N,d] @ LyV):
    ti, tj = dato.get_tid()
    o[ti,tj].put(dato.matmul(w[ti,tj].get(),V))

  @dato.task(mapping=[P0,1])
  def acc(O: Ty[N,d] @ LyQ):
    output: Ty[Tq,d] = 0
    ti, _ = dato.get_tid()
    for i in range(P1):
      rescale(output, exp_scale[ti].get())
      output[:,:] = dato.add(
        output, o[ti,i].get())
    scale(output,exp_sum[ti].get(),O)
\end{minted}
\end{minipage}
% \caption{Flash Attention kernel in \Name}
% \label{subfig:mha_code}
\end{subfigure}
\begin{subfigure}[b]{0.35\linewidth}
\centering
\includegraphics[width=\linewidth]{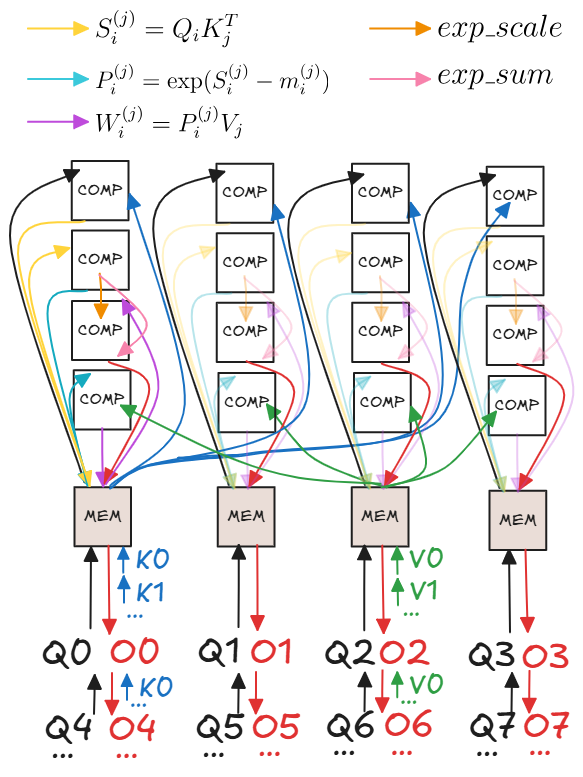}
% \caption{Physical mapping of the attention kernel}
% \label{subfig:mha_mapping}
\end{subfigure}
% \begin{subfigure}[b]{0.3\linewidth}
% \centering
% \caption{MHA Trace}
% \label{subfig:trace}
% \end{subfigure}
% \hfill
\caption{\textbf{Left:} Implementation of the Flash Attention kernel in \Name. \textbf{Right:} Physical mapping of the attention kernel.}
\label{fig:mha}
% required by acmart, for voiceover to read the paper aloud
\Description{}
\end{figure*}

\textbf{Multi-Head Attention (MHA).}
To fully exploit the advantage of the dataflow architecture, we implement the FlashAttention kernel, a fused MHA using online softmax, on the NPU, enabling streaming MHA without returns to DRAM. Fig.~\ref{fig:mha} shows the \Name implementation and the corresponding physical mapping, where the stream and layout types make the program easier to write and read. As reported in Fig.~\ref{fig:mha_ffn_result}, \Name consistently outperforms IRON across sequence lengths, with a maximum speedup of 2.81$\times$. The gains are due to: (i) reduced off-chip traffic via on-chip streaming and spatial scaling, which eliminates write-back/read-back between kernels; (ii) lower kernel-launch overhead by running a single fused pipeline across tiles; and (iii) a smaller per-tile working set that allows \Name to exceed sequence length 2048, where IRON fails due to DMA transfer limitations.
% \fixme{} \zz{move it to a more prounced location, maybe in the leading paragraph of 7.2.2? also do we want to add FlashAttention to the title of the paragraph?}

\textbf{Feed-Forward Network (FFN).}
In \Name, we spatially fuse the two GEMMs into a single dataflow pipeline, partitioning the $4\times 4$ array into two $4\times 2$ submeshes for each GEMM. In contrast, IRON cannot fuse these kernels: it invokes two standalone GEMMs sequentially, forcing global-memory write-back/read-back between stages and incurring per-kernel launch overheads. As shown in Fig.~\ref{fig:mha_ffn_result}, this fused execution yields a 1.64$\times$ speedup over IRON. The gains stem from reduced off-chip traffic, improved pipeline occupancy, and better tile utilization.

\subsection{Scalability}
% \begin{figure}[!htbp]
% \includegraphics[width=0.8\linewidth]{figures/gemm-ablation.pdf}
% \caption{Ablation study on different optimizations.}
% \label{fig:ablation}
% % required by acmart, for voiceover to read the paper aloud
% \Description{}
% \end{figure}

\begin{figure}[t]
\begin{tabular}{cc}
\includegraphics[width=0.335\linewidth]{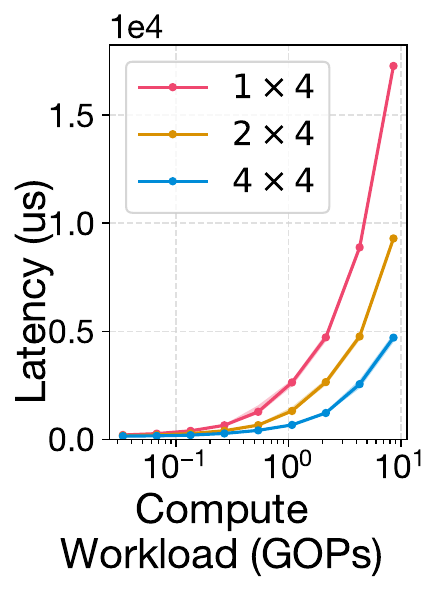} &
\includegraphics[width=0.65\linewidth]{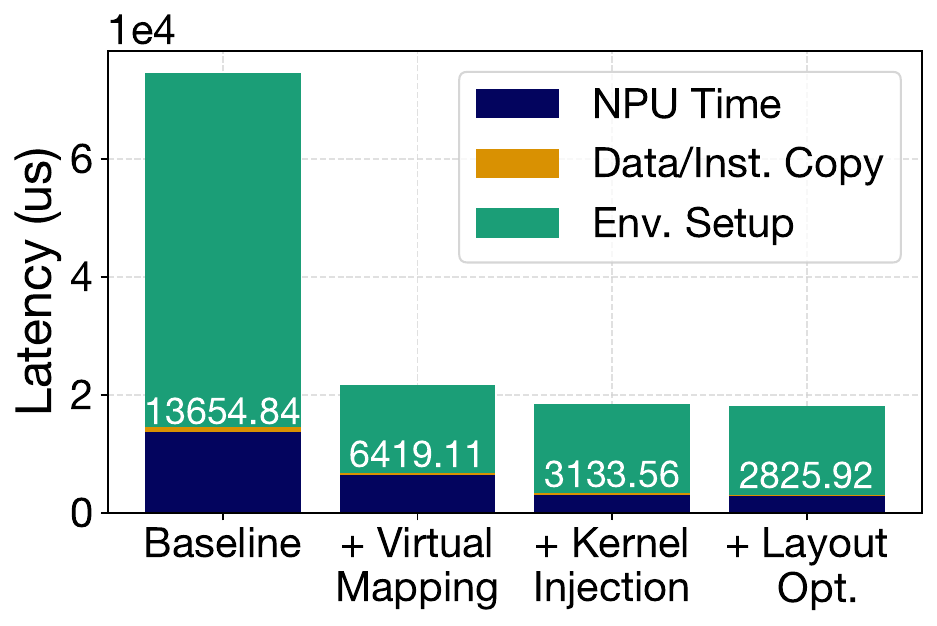}
\end{tabular}
\caption{\textbf{Left:} Scalability results on GEMM. \textbf{Right:} Optimization ablation study w/ GEMM. The numbers on the bar denote NPU time.}
\label{fig:ablation-scalability}
\Description{}
\end{figure}
Fig.~\ref{fig:ablation-scalability} reports GEMM scalability on the NPU as the active mesh increases from $1\times4$ to $2\times4$ and the full $4\times4$ array. For the largest workload, absolute latency decreases nearly proportionally with tile count: the $2\times4$ configuration achieves a 1.97$\times$ speedup over $1\times4$, while the $4\times4$ configuration reaches 3.67$\times$. Although utilization declines slightly due to growing communication and reduction overheads, the results demonstrate that \Name scales effectively with the number of compute tiles, highlighting the importance of its virtual mapping, layout optimization, and DMA scheduling in sustaining utilization at scale.

\subsection{Ablation Study}
\label{sub:ablation}
Fig.~\ref{fig:ablation-scalability} presents an ablation of \Name's optimization stack, decomposing latency into NPU compute, data/instruction copies, and environment setup. Starting from the baseline, NPU time is \SI{13.65}{ms}; enabling virtual mapping reduces it to \SI{6.42}{ms} while also collapsing the dominant setup overhead, yielding the largest end-to-end gain. Injecting vectorized kernels halves the compute time to \SI{3.13}{ms} (–51\%), and subsequent layout optimization trims it further to \SI{2.83}{ms} (–10\%). Across the sequence, data/instruction copy costs remain negligible, and the total latency drops by roughly $4\times$ relative to the baseline. These results indicate that virtual mapping primarily removes orchestration and launch costs, and other optimizations drive the remaining improvements by increasing tile utilization and reducing transformation overheads.

\begin{figure}[t]
\begin{tabular}{cc}
\includegraphics[width=0.48\linewidth]{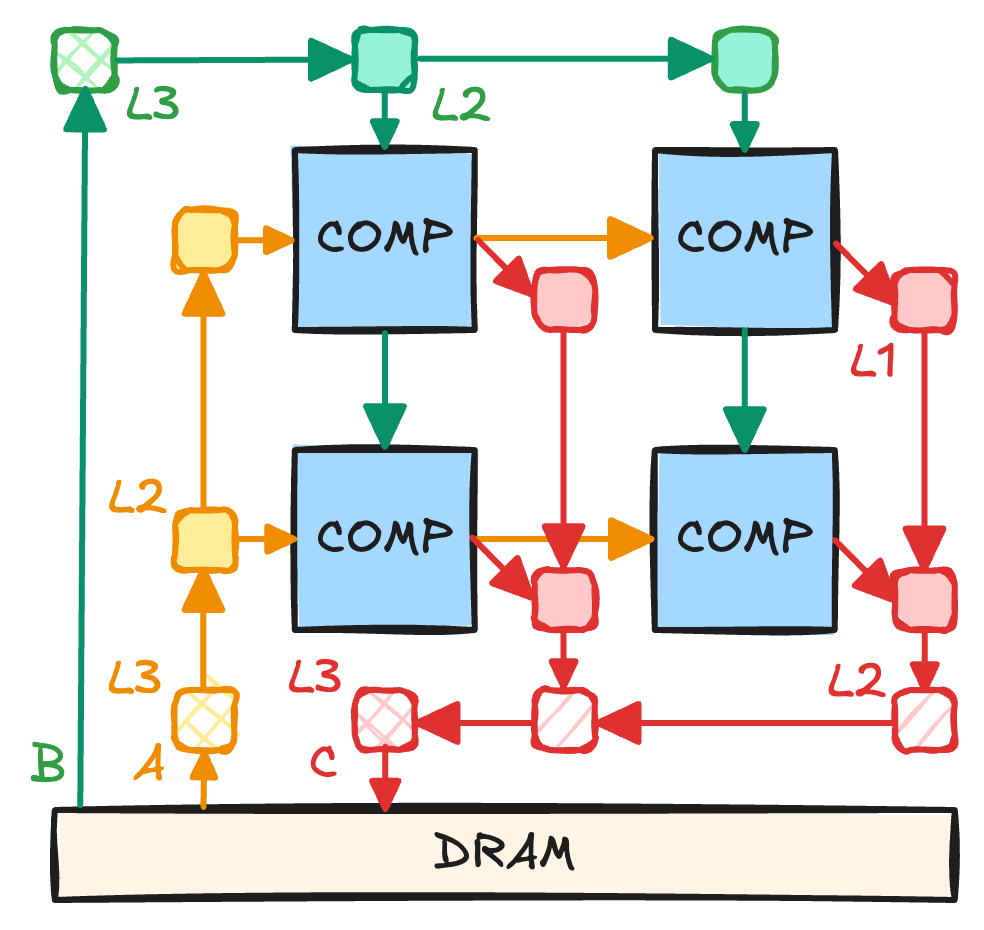} &
\includegraphics[width=0.52\linewidth]{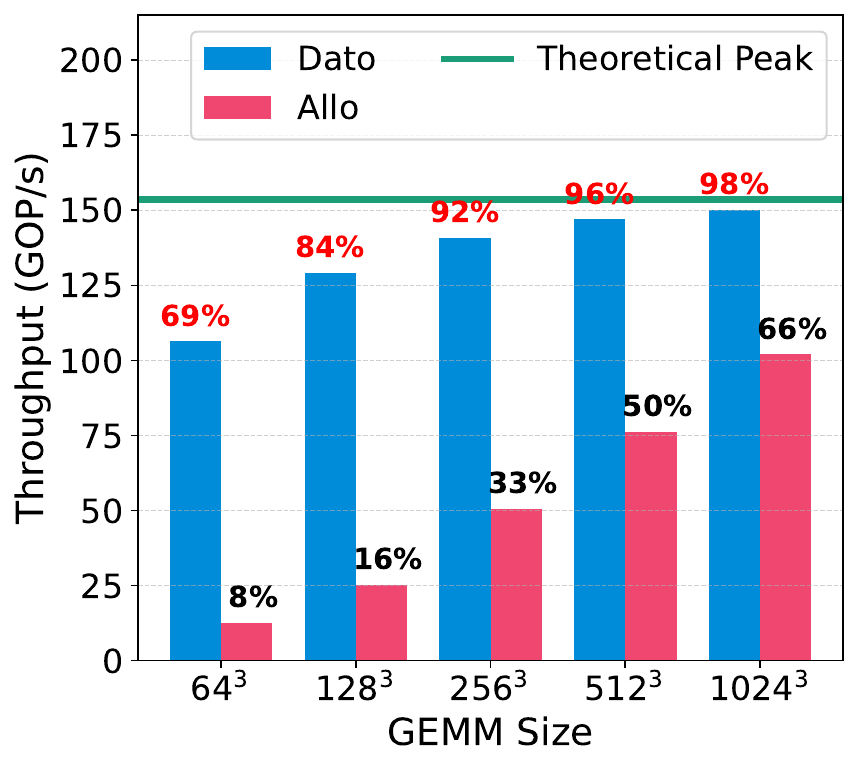}
\end{tabular}
\caption{\textbf{Left:} \Name-generated systolic array architecture. \textbf{Right:} Performance results on Alveo U280 FPGA. The labeled numbers are the achieved hardware utilization.}
\label{fig:fpga}
% required by acmart, for voiceover to read the paper aloud
\Description{}
\end{figure}

\subsection{Portability}
% \zz{we need to elaborate the discussion on FPGA -- reiterate why we add this target; if and how the program differs from the one for NPU and which compilation steps are different}
To demonstrate \Name's portability beyond NPUs, we generate a $16\times 16$ \texttt{i8} output-stationary systolic array running at \SI{300}{MHz} targeting AMD Alveo U280 FPGA, a drastically different substrate.
Users can retain the same \Name frontend and simply specify the FPGA as the build target.
We implement a daisy-chain systolic array architecture as shown in Fig.~\ref{fig:fpga}, where each PE and the multi-level caches are encapsulated as \mintinline{python}{dato.task} modules and interconnected using \Name's streams.
The full implementation is provided in Appendix~\ref{appendix:sa}. For this backend, the compiler bypasses NPU-specific DMA optimizations and instead generates C++ HLS code for FPGA synthesis.

We evaluate GEMM workloads ranging from $64^3$ to $1024^3$ and compare against the state-of-the-art framework Allo~\cite{chen2024allo}. Fig.~\ref{fig:fpga} shows \Name consistently outperforms Allo by up to $8.2\times$. At the largest matrix size, \Name achieves \SI{150}{GOP/s}, approaching the systolic array's theoretical peak of \SI{153.6}{GOP/s}, assuming one MAC per cycle for each PE.
The key to this performance is \Name's ability to achieve higher PE utilization by overlapping data transfer and computation through a carefully designed multi-level memory hierarchy, a transformation that Allo cannot realize from a simple GEMM kernel given its limited set of primitives.
Moreover, the results after placement and routing show \Name reduces output fan-out pressure and enables the design to meet the \SI{300}{MHz} timing target. In contrast, Allo fails to meet timing, achieving only \SI{132}{MHz} due to excessive fan-out.

\begin{table}[t]
\caption{Line of code comparison on different kernels.}
\label{tab:loc}
\resizebox{0.9\linewidth}{!}{
\begin{tabular}{cccc}\Xhline{2\arrayrulewidth}
\textbf{Framework} & \textbf{GEMM} & \textbf{MHA} & \textbf{FFN}\\\hline
IRON/\Name & 101/8 (12$\times$) & 239/44 (5$\times$) & 101/14 (7$\times$)\\\hline
% Reduction & 10$\times$ & 5$\times$ & 7$\times$\\\hline
\Xhline{2\arrayrulewidth}
\end{tabular}
}
\end{table}

% XDNA2 results

\subsection{Productivity}
Finally, we compare implementation effort in terms of lines of code (LoC). As shown in Table~\ref{tab:loc}, \Name matches the performance of IRON while using 12$\times$ fewer lines of code (see examples in Fig.~\ref{subfig:tiled_gemm} and~\ref{fig:mha}). Beyond conciseness, \Name also unlocks spatial multi-kernel fusion and seamless retargeting to other accelerators---capabilities that significantly enhance the productivity of performance engineers.

% compilation time
% power efficiency
\section{Conclusion and Future Work}
\label{sec:conclusion}
In this paper, we introduce \Name, a Python-embedded, task-based programming model that elevates data communication and sharding to first-class types and compiles virtual mappings to efficient physical deployments on dataflow accelerators.
As future work, we will extend the compiler from local decisions to global graph-level optimization. We also plan to automatically generate high-performance kernels from task specifications, reducing reliance on hand-written libraries. Finally, we will broaden targets and evaluation to accelerators such as NVIDIA Hopper and successors with TMA, as well as other programmable dataflow fabrics.

% use the ACM bibliography style
\newpage
\bibliographystyle{ACM-Reference-Format}
\bibliography{asplos26}

%%% -*-BibTeX-*-
%%% Do NOT edit. File created by BibTeX with style
%%% ACM-Reference-Format-Journals [18-Jan-2012].

\begin{thebibliography}{62}

%%% ====================================================================
%%% NOTE TO THE USER: you can override these defaults by providing
%%% customized versions of any of these macros before the \bibliography
%%% command.  Each of them MUST provide its own final punctuation,
%%% except for \shownote{} and \showURL{}.  The latter two
%%% do not use final punctuation, in order to avoid confusing it with
%%% the Web address.
%%%
%%% To suppress output of a particular field, define its macro to expand
%%% to an empty string, or better, \unskip, like this:
%%%
%%% \newcommand{\showURL}[1]{\unskip}   % LaTeX syntax
%%%
%%% \def \showURL #1{\unskip}           % plain TeX syntax
%%%
%%% ====================================================================

\ifx \showCODEN    \undefined \def \showCODEN     #1{\unskip}     \fi
\ifx \showISBNx    \undefined \def \showISBNx     #1{\unskip}     \fi
\ifx \showISBNxiii \undefined \def \showISBNxiii  #1{\unskip}     \fi
\ifx \showISSN     \undefined \def \showISSN      #1{\unskip}     \fi
\ifx \showLCCN     \undefined \def \showLCCN      #1{\unskip}     \fi
\ifx \shownote     \undefined \def \shownote      #1{#1}          \fi
\ifx \showarticletitle \undefined \def \showarticletitle #1{#1}   \fi
\ifx \showURL      \undefined \def \showURL       {\relax}        \fi
% The following commands are used for tagged output and should be
% invisible to TeX
\providecommand\bibfield[2]{#2}
\providecommand\bibinfo[2]{#2}
\providecommand\natexlab[1]{#1}
\providecommand\showeprint[2][]{arXiv:#2}

\bibitem[Alabed et~al\mbox{.}(2025)]%
        {sami2025partir}
\bibfield{author}{\bibinfo{person}{Sami Alabed}, \bibinfo{person}{Daniel Belov}, \bibinfo{person}{Bart Chrzaszcz}, \bibinfo{person}{Juliana Franco}, \bibinfo{person}{Dominik Grewe}, \bibinfo{person}{Dougal Maclaurin}, \bibinfo{person}{James Molloy}, \bibinfo{person}{Tom Natan}, \bibinfo{person}{Tamara Norman}, \bibinfo{person}{Xiaoyue Pan}, \bibinfo{person}{Adam Paszke}, \bibinfo{person}{Norman~A. Rink}, \bibinfo{person}{Michael Schaarschmidt}, \bibinfo{person}{Timur Sitdikov}, \bibinfo{person}{Agnieszka Swietlik}, \bibinfo{person}{Dimitrios Vytiniotis}, {and} \bibinfo{person}{Joel Wee}.} \bibinfo{year}{2025}\natexlab{}.
\newblock \showarticletitle{PartIR: Composing SPMD Partitioning Strategies for Machine Learning}. In \bibinfo{booktitle}{\emph{Proceedings of the 30th ACM International Conference on Architectural Support for Programming Languages and Operating Systems, Volume 1}} (Rotterdam, Netherlands) \emph{(\bibinfo{series}{ASPLOS '25})}. \bibinfo{publisher}{Association for Computing Machinery}, \bibinfo{address}{New York, NY, USA}, \bibinfo{pages}{794–810}.
\newblock
\showISBNx{9798400706981}
\href{https://doi.org/10.1145/3669940.3707284}{doi:\nolinkurl{10.1145/3669940.3707284}}


\bibitem[AWS(2023)]%
        {aws_inferentia}
\bibfield{author}{\bibinfo{person}{AWS}.} \bibinfo{year}{2023}\natexlab{}.
\newblock \bibinfo{title}{Inferentia Architecture}.
\newblock \bibinfo{howpublished}{\url{https://awsdocs-neuron.readthedocs-hosted.com/en/latest/general/arch/neuron-hardware/inferentia.html}}.
\newblock


\bibitem[Basalama and Cong(2025)]%
        {suhail2025streamhls}
\bibfield{author}{\bibinfo{person}{Suhail Basalama} {and} \bibinfo{person}{Jason Cong}.} \bibinfo{year}{2025}\natexlab{}.
\newblock \showarticletitle{Stream-HLS: Towards Automatic Dataflow Acceleration}. In \bibinfo{booktitle}{\emph{Proceedings of the 2025 ACM/SIGDA International Symposium on Field Programmable Gate Arrays}}. \bibinfo{publisher}{ACM}, \bibinfo{address}{New York, NY, USA}, \bibinfo{pages}{103--114}.
\newblock


\bibitem[Cerebras(2024)]%
        {cerebras_wse}
\bibfield{author}{\bibinfo{person}{Cerebras}.} \bibinfo{year}{2024}\natexlab{}.
\newblock \bibinfo{title}{The Future of AI is Wafer Scale}.
\newblock \bibinfo{howpublished}{\url{https://www.cerebras.ai/chip}}.
\newblock


\bibitem[Chen et~al\mbox{.}(2024a)]%
        {chen2024slapo}
\bibfield{author}{\bibinfo{person}{Hongzheng Chen}, \bibinfo{person}{Cody~Hao Yu}, \bibinfo{person}{Shuai Zheng}, \bibinfo{person}{Zhen Zhang}, \bibinfo{person}{Zhiru Zhang}, {and} \bibinfo{person}{Yida Wang}.} \bibinfo{year}{2024}\natexlab{a}.
\newblock \showarticletitle{Slapo: A Schedule Language for Progressive Optimization of Large Deep Learning Model Training}. In \bibinfo{booktitle}{\emph{Proceedings of the 29th ACM International Conference on Architectural Support for Programming Languages and Operating Systems, Volume 2}} (La Jolla, CA, USA) \emph{(\bibinfo{series}{ASPLOS'24})}. \bibinfo{publisher}{Association for Computing Machinery}, \bibinfo{address}{New York, NY, USA}, \bibinfo{pages}{1095--1111}.
\newblock


\bibitem[Chen et~al\mbox{.}(2024b)]%
        {chen2024understanding}
\bibfield{author}{\bibinfo{person}{Hongzheng Chen}, \bibinfo{person}{Jiahao Zhang}, \bibinfo{person}{Yixiao Du}, \bibinfo{person}{Shaojie Xiang}, \bibinfo{person}{Zichao Yue}, \bibinfo{person}{Niansong Zhang}, \bibinfo{person}{Yaohui Cai}, {and} \bibinfo{person}{Zhiru Zhang}.} \bibinfo{year}{2024}\natexlab{b}.
\newblock \showarticletitle{Understanding the potential of fpga-based spatial acceleration for large language model inference}.
\newblock \bibinfo{journal}{\emph{ACM Transactions on Reconfigurable Technology and Systems}} \bibinfo{volume}{18}, \bibinfo{number}{1} (\bibinfo{year}{2024}), \bibinfo{pages}{1--29}.
\newblock


\bibitem[Chen et~al\mbox{.}(2024c)]%
        {chen2024allo}
\bibfield{author}{\bibinfo{person}{Hongzheng Chen}, \bibinfo{person}{Niansong Zhang}, \bibinfo{person}{Shaojie Xiang}, \bibinfo{person}{Zhichen Zeng}, \bibinfo{person}{Mengjia Dai}, {and} \bibinfo{person}{Zhiru Zhang}.} \bibinfo{year}{2024}\natexlab{c}.
\newblock \showarticletitle{Allo: A programming model for composable accelerator design}.
\newblock \bibinfo{journal}{\emph{Proceedings of the ACM on Programming Languages}} \bibinfo{volume}{8}, \bibinfo{number}{PLDI} (\bibinfo{year}{2024}), \bibinfo{pages}{593--620}.
\newblock


\bibitem[Chen et~al\mbox{.}(2018)]%
        {chen2018tvm}
\bibfield{author}{\bibinfo{person}{Tianqi Chen}, \bibinfo{person}{Thierry Moreau}, \bibinfo{person}{Ziheng Jiang}, \bibinfo{person}{Lianmin Zheng}, \bibinfo{person}{Eddie Yan}, \bibinfo{person}{Meghan Cowan}, \bibinfo{person}{Haichen Shen}, \bibinfo{person}{Leyuan Wang}, \bibinfo{person}{Yuwei Hu}, \bibinfo{person}{Luis Ceze}, \bibinfo{person}{Carlos Guestrin}, {and} \bibinfo{person}{Arvind Krishnamurthy}.} \bibinfo{year}{2018}\natexlab{}.
\newblock \showarticletitle{TVM: An Automated End-to-End Optimizing Compiler for Deep Learning}. In \bibinfo{booktitle}{\emph{Proceedings of the 13th USENIX Conference on Operating Systems Design and Implementation}} (Carlsbad, CA, USA) \emph{(\bibinfo{series}{OSDI'18})}. \bibinfo{publisher}{USENIX Association}, \bibinfo{address}{USA}, \bibinfo{pages}{579–594}.
\newblock
\showISBNx{9781931971478}


\bibitem[Cutler et~al\mbox{.}(2024)]%
        {cutler2024streamtype}
\bibfield{author}{\bibinfo{person}{Joseph~W Cutler}, \bibinfo{person}{Christopher Watson}, \bibinfo{person}{Emeka Nkurumeh}, \bibinfo{person}{Phillip Hilliard}, \bibinfo{person}{Harrison Goldstein}, \bibinfo{person}{Caleb Stanford}, {and} \bibinfo{person}{Benjamin~C Pierce}.} \bibinfo{year}{2024}\natexlab{}.
\newblock \showarticletitle{Stream types}.
\newblock \bibinfo{journal}{\emph{Proceedings of the ACM on Programming Languages}} \bibinfo{volume}{8}, \bibinfo{number}{PLDI} (\bibinfo{year}{2024}), \bibinfo{pages}{1412--1436}.
\newblock


\bibitem[Dadu and Nowatzki(2022)]%
        {dadu2022taskstream}
\bibfield{author}{\bibinfo{person}{Vidushi Dadu} {and} \bibinfo{person}{Tony Nowatzki}.} \bibinfo{year}{2022}\natexlab{}.
\newblock \showarticletitle{TaskStream: accelerating task-parallel workloads by recovering program structure}. In \bibinfo{booktitle}{\emph{Proceedings of the 27th ACM International Conference on Architectural Support for Programming Languages and Operating Systems}} (Lausanne, Switzerland) \emph{(\bibinfo{series}{ASPLOS '22})}. \bibinfo{publisher}{Association for Computing Machinery}, \bibinfo{address}{New York, NY, USA}, \bibinfo{pages}{1–13}.
\newblock
\showISBNx{9781450392051}
\href{https://doi.org/10.1145/3503222.3507706}{doi:\nolinkurl{10.1145/3503222.3507706}}


\bibitem[Dao et~al\mbox{.}(2022)]%
        {dao2022flashattention}
\bibfield{author}{\bibinfo{person}{Tri Dao}, \bibinfo{person}{Dan Fu}, \bibinfo{person}{Stefano Ermon}, \bibinfo{person}{Atri Rudra}, {and} \bibinfo{person}{Christopher R{\'e}}.} \bibinfo{year}{2022}\natexlab{}.
\newblock \showarticletitle{Flashattention: Fast and memory-efficient exact attention with io-awareness}.
\newblock \bibinfo{journal}{\emph{Advances in neural information processing systems}}  \bibinfo{volume}{35} (\bibinfo{year}{2022}), \bibinfo{pages}{16344--16359}.
\newblock


\bibitem[Ding et~al\mbox{.}(2025)]%
        {ding2025tilus}
\bibfield{author}{\bibinfo{person}{Yaoyao Ding}, \bibinfo{person}{Bohan Hou}, \bibinfo{person}{Xiao Zhang}, \bibinfo{person}{Allan Lin}, \bibinfo{person}{Tianqi Chen}, \bibinfo{person}{Cody~Yu Hao}, \bibinfo{person}{Yida Wang}, {and} \bibinfo{person}{Gennady Pekhimenko}.} \bibinfo{year}{2025}\natexlab{}.
\newblock \bibinfo{title}{Tilus: A Tile-Level GPGPU Programming Language for Low-Precision Computation}.
\newblock
\showeprint[arxiv]{2504.12984}~[cs.LG]
\urldef\tempurl%
\url{https://arxiv.org/abs/2504.12984}
\showURL{%
\tempurl}


\bibitem[Dubey et~al\mbox{.}(2024)]%
        {dubey2024llama3}
\bibfield{author}{\bibinfo{person}{Abhimanyu Dubey}, \bibinfo{person}{Abhinav Jauhri}, \bibinfo{person}{Abhinav Pandey}, \bibinfo{person}{Abhishek Kadian}, \bibinfo{person}{Ahmad Al-Dahle}, \bibinfo{person}{Aiesha Letman}, \bibinfo{person}{Akhil Mathur}, \bibinfo{person}{Alan Schelten}, \bibinfo{person}{Amy Yang}, \bibinfo{person}{Angela Fan}, {et~al\mbox{.}}} \bibinfo{year}{2024}\natexlab{}.
\newblock \bibinfo{title}{The Llama 3 Herd of Models}.
\newblock
\showeprint[arxiv]{2407.21783}~[cs.AI]
\urldef\tempurl%
\url{https://arxiv.org/abs/2407.21783}
\showURL{%
\tempurl}


\bibitem[Durst et~al\mbox{.}(2020)]%
        {durst2020aetherling}
\bibfield{author}{\bibinfo{person}{David Durst}, \bibinfo{person}{Matthew Feldman}, \bibinfo{person}{Dillon Huff}, \bibinfo{person}{David Akeley}, \bibinfo{person}{Ross Daly}, \bibinfo{person}{Gilbert~Louis Bernstein}, \bibinfo{person}{Marco Patrignani}, \bibinfo{person}{Kayvon Fatahalian}, {and} \bibinfo{person}{Pat Hanrahan}.} \bibinfo{year}{2020}\natexlab{}.
\newblock \showarticletitle{Type-Directed Scheduling of Streaming Accelerators}. In \bibinfo{booktitle}{\emph{Proceedings of the 41st ACM SIGPLAN Conference on Programming Language Design and Implementation}} (London, UK) \emph{(\bibinfo{series}{PLDI 2020})}. \bibinfo{publisher}{Association for Computing Machinery}, \bibinfo{address}{New York, NY, USA}, \bibinfo{pages}{408–422}.
\newblock
\showISBNx{9781450376136}
\href{https://doi.org/10.1145/3385412.3385983}{doi:\nolinkurl{10.1145/3385412.3385983}}


\bibitem[Emani et~al\mbox{.}(2021)]%
        {emani2021sambanova}
\bibfield{author}{\bibinfo{person}{Murali Emani}, \bibinfo{person}{Venkatram Vishwanath}, \bibinfo{person}{Corey Adams}, \bibinfo{person}{Michael~E. Papka}, \bibinfo{person}{Rick Stevens}, \bibinfo{person}{Laura Florescu}, \bibinfo{person}{Sumti Jairath}, \bibinfo{person}{William Liu}, \bibinfo{person}{Tejas Nama}, {and} \bibinfo{person}{Arvind Sujeeth}.} \bibinfo{year}{2021}\natexlab{}.
\newblock \showarticletitle{Accelerating Scientific Applications With SambaNova Reconfigurable Dataflow Architecture}.
\newblock \bibinfo{journal}{\emph{Computing in Science \& Engineering}} \bibinfo{volume}{23}, \bibinfo{number}{2} (\bibinfo{year}{2021}), \bibinfo{pages}{114--119}.
\newblock
\href{https://doi.org/10.1109/MCSE.2021.3057203}{doi:\nolinkurl{10.1109/MCSE.2021.3057203}}


\bibitem[Fu et~al\mbox{.}(2024)]%
        {fu2024trainium}
\bibfield{author}{\bibinfo{person}{Xinwei Fu}, \bibinfo{person}{Zhen Zhang}, \bibinfo{person}{Haozheng Fan}, \bibinfo{person}{Guangtai Huang}, \bibinfo{person}{Mohammad El-Shabani}, \bibinfo{person}{Randy Huang}, \bibinfo{person}{Rahul Solanki}, \bibinfo{person}{Fei Wu}, \bibinfo{person}{Ron Diamant}, {and} \bibinfo{person}{Yida Wang}.} \bibinfo{year}{2024}\natexlab{}.
\newblock \showarticletitle{Distributed Training of Large Language Models on AWS Trainium}. In \bibinfo{booktitle}{\emph{Proceedings of the 2024 ACM Symposium on Cloud Computing}} (Redmond, WA, USA) \emph{(\bibinfo{series}{SoCC '24})}. \bibinfo{publisher}{Association for Computing Machinery}, \bibinfo{address}{New York, NY, USA}, \bibinfo{pages}{961–976}.
\newblock
\showISBNx{9798400712869}
\href{https://doi.org/10.1145/3698038.3698535}{doi:\nolinkurl{10.1145/3698038.3698535}}


\bibitem[Gholami et~al\mbox{.}(2024)]%
        {amir2024memorywall}
\bibfield{author}{\bibinfo{person}{Amir Gholami}, \bibinfo{person}{Zhewei Yao}, \bibinfo{person}{Sehoon Kim}, \bibinfo{person}{Coleman Hooper}, \bibinfo{person}{Michael~W Mahoney}, {and} \bibinfo{person}{Kurt Keutzer}.} \bibinfo{year}{2024}\natexlab{}.
\newblock \showarticletitle{Ai and memory wall}.
\newblock \bibinfo{journal}{\emph{IEEE Micro}} \bibinfo{volume}{44}, \bibinfo{number}{3} (\bibinfo{year}{2024}), \bibinfo{pages}{33--39}.
\newblock


\bibitem[Ghosh et~al\mbox{.}(2025)]%
        {ghosh2025ripple}
\bibfield{author}{\bibinfo{person}{Souradip Ghosh}, \bibinfo{person}{Yufei Shi}, \bibinfo{person}{Brandon Lucia}, {and} \bibinfo{person}{Nathan Beckmann}.} \bibinfo{year}{2025}\natexlab{}.
\newblock \showarticletitle{Ripple: Asynchronous Programming for Spatial Dataflow Architectures}.
\newblock \bibinfo{journal}{\emph{Proc. ACM Program. Lang.}} \bibinfo{volume}{9}, \bibinfo{number}{PLDI}, Article \bibinfo{articleno}{157} (\bibinfo{date}{June} \bibinfo{year}{2025}), \bibinfo{numpages}{28}~pages.
\newblock
\href{https://doi.org/10.1145/3729256}{doi:\nolinkurl{10.1145/3729256}}


\bibitem[Google(2025)]%
        {tpuv7}
\bibfield{author}{\bibinfo{person}{Google}.} \bibinfo{year}{2025}\natexlab{}.
\newblock \bibinfo{title}{Ironwood: The first Google TPU for the age of inference}.
\newblock \bibinfo{howpublished}{\url{https://blog.google/products/google-cloud/ironwood-tpu-age-of-inference/}}.
\newblock


\bibitem[Guo et~al\mbox{.}(2023)]%
        {guo2023tapa}
\bibfield{author}{\bibinfo{person}{Licheng Guo}, \bibinfo{person}{Yuze Chi}, \bibinfo{person}{Jason Lau}, \bibinfo{person}{Linghao Song}, \bibinfo{person}{Xingyu Tian}, \bibinfo{person}{Moazin Khatti}, \bibinfo{person}{Weikang Qiao}, \bibinfo{person}{Jie Wang}, \bibinfo{person}{Ecenur Ustun}, \bibinfo{person}{Zhenman Fang}, \bibinfo{person}{Zhiru Zhang}, {and} \bibinfo{person}{Jason Cong}.} \bibinfo{year}{2023}\natexlab{}.
\newblock \showarticletitle{TAPA: A Scalable Task-Parallel Dataflow Programming Framework for Modern FPGAs with Co-Optimization of HLS and Physical Design}.
\newblock \bibinfo{journal}{\emph{ACM Trans. Reconfigurable Technol. Syst.}} \bibinfo{volume}{16}, \bibinfo{number}{4} (\bibinfo{date}{sep} \bibinfo{year}{2023}), \bibinfo{pages}{1--31}.
\newblock
\showISSN{1936-7406}
\href{https://doi.org/10.1145/3609335}{doi:\nolinkurl{10.1145/3609335}}


\bibitem[Hagedorn et~al\mbox{.}(2023)]%
        {hagedorn2023graphene}
\bibfield{author}{\bibinfo{person}{Bastian Hagedorn}, \bibinfo{person}{Bin Fan}, \bibinfo{person}{Hanfeng Chen}, \bibinfo{person}{Cris Cecka}, \bibinfo{person}{Michael Garland}, {and} \bibinfo{person}{Vinod Grover}.} \bibinfo{year}{2023}\natexlab{}.
\newblock \showarticletitle{Graphene: An ir for optimized tensor computations on gpus}. In \bibinfo{booktitle}{\emph{Proceedings of the 28th ACM International Conference on Architectural Support for Programming Languages and Operating Systems, Volume 3}}. \bibinfo{publisher}{Association for Computing Machinery}, \bibinfo{address}{New York, NY, USA}, \bibinfo{pages}{302--313}.
\newblock


\bibitem[Huang et~al\mbox{.}(2022)]%
        {huangtaskflow2022}
\bibfield{author}{\bibinfo{person}{Tsung-Wei Huang}, \bibinfo{person}{Dian-Lun Lin}, \bibinfo{person}{Chun-Xun Lin}, {and} \bibinfo{person}{Yibo Lin}.} \bibinfo{year}{2022}\natexlab{}.
\newblock \showarticletitle{Taskflow: A Lightweight Parallel and Heterogeneous Task Graph Computing System}.
\newblock \bibinfo{journal}{\emph{IEEE Transactions on Parallel and Distributed Systems}} \bibinfo{volume}{33}, \bibinfo{number}{6} (\bibinfo{year}{2022}), \bibinfo{pages}{1303--1320}.
\newblock
\href{https://doi.org/10.1109/TPDS.2021.3104255}{doi:\nolinkurl{10.1109/TPDS.2021.3104255}}


\bibitem[Hunhoff et~al\mbox{.}(2025)]%
        {hunhoff2025iron}
\bibfield{author}{\bibinfo{person}{Erika Hunhoff}, \bibinfo{person}{Joseph Melber}, \bibinfo{person}{Kristof Denolf}, \bibinfo{person}{Andra Bisca}, \bibinfo{person}{Samuel Bayliss}, \bibinfo{person}{Stephen Neuendorffer}, \bibinfo{person}{Jeff Fifield}, \bibinfo{person}{Jack Lo}, \bibinfo{person}{Pranathi Vasireddy}, \bibinfo{person}{Phil James-Roxby}, {et~al\mbox{.}}} \bibinfo{year}{2025}\natexlab{}.
\newblock \showarticletitle{Efficiency, Expressivity, and Extensibility in a Close-to-Metal NPU Programming Interface}. In \bibinfo{booktitle}{\emph{2025 IEEE 33rd Annual International Symposium on Field-Programmable Custom Computing Machines (FCCM)}}. \bibinfo{publisher}{IEEE}, \bibinfo{address}{Los Alamitos, CA, USA}, \bibinfo{pages}{85--94}.
\newblock


\bibitem[IBM(2024)]%
        {ibm_spyre}
\bibfield{author}{\bibinfo{person}{IBM}.} \bibinfo{year}{2024}\natexlab{}.
\newblock \bibinfo{title}{Enhancing enterprise AI with the IBM Spyre Accelerator}.
\newblock \bibinfo{howpublished}{\url{https://research.ibm.com/blog/spyre-for-z}}.
\newblock


\bibitem[IBM(2025)]%
        {vonneumann_bottleneck}
\bibfield{author}{\bibinfo{person}{IBM}.} \bibinfo{year}{2025}\natexlab{}.
\newblock \bibinfo{title}{Why a decades old architecture decision is impeding the power of AI computing}.
\newblock \bibinfo{howpublished}{\url{https://research.ibm.com/blog/why-von-neumann-architecture-is-impeding-the-power-of-ai-computing}}.
\newblock


\bibitem[Ikarashi et~al\mbox{.}(2022)]%
        {yuka2022exo}
\bibfield{author}{\bibinfo{person}{Yuka Ikarashi}, \bibinfo{person}{Gilbert~Louis Bernstein}, \bibinfo{person}{Alex Reinking}, \bibinfo{person}{Hasan Genc}, {and} \bibinfo{person}{Jonathan Ragan-Kelley}.} \bibinfo{year}{2022}\natexlab{}.
\newblock \showarticletitle{Exocompilation for Productive Programming of Hardware Accelerators}. In \bibinfo{booktitle}{\emph{Proceedings of the 43rd ACM SIGPLAN International Conference on Programming Language Design and Implementation}} (San Diego, CA, USA) \emph{(\bibinfo{series}{PLDI 2022})}. \bibinfo{publisher}{Association for Computing Machinery}, \bibinfo{address}{New York, NY, USA}, \bibinfo{pages}{703–718}.
\newblock
\showISBNx{9781450392655}
\href{https://doi.org/10.1145/3519939.3523446}{doi:\nolinkurl{10.1145/3519939.3523446}}


\bibitem[Ikarashi et~al\mbox{.}(2025)]%
        {ikarashi2025exo2}
\bibfield{author}{\bibinfo{person}{Yuka Ikarashi}, \bibinfo{person}{Kevin Qian}, \bibinfo{person}{Samir Droubi}, \bibinfo{person}{Alex Reinking}, \bibinfo{person}{Gilbert~Louis Bernstein}, {and} \bibinfo{person}{Jonathan Ragan-Kelley}.} \bibinfo{year}{2025}\natexlab{}.
\newblock \showarticletitle{Exo 2: Growing a Scheduling Language}. In \bibinfo{booktitle}{\emph{Proceedings of the 30th ACM International Conference on Architectural Support for Programming Languages and Operating Systems, Volume 1}}. \bibinfo{publisher}{ACM}, \bibinfo{address}{New York, NY, USA}, \bibinfo{pages}{426--444}.
\newblock


\bibitem[Jouppi et~al\mbox{.}(2023)]%
        {jouppi2023tpuv4}
\bibfield{author}{\bibinfo{person}{Norm Jouppi}, \bibinfo{person}{George Kurian}, \bibinfo{person}{Sheng Li}, \bibinfo{person}{Peter Ma}, \bibinfo{person}{Rahul Nagarajan}, {et~al\mbox{.}}} \bibinfo{year}{2023}\natexlab{}.
\newblock \showarticletitle{TPU v4: An Optically Reconfigurable Supercomputer for Machine Learning with Hardware Support for Embeddings}. In \bibinfo{booktitle}{\emph{Proceedings of the 50th Annual International Symposium on Computer Architecture}} (Orlando, FL, USA) \emph{(\bibinfo{series}{ISCA'23})}. \bibinfo{publisher}{Association for Computing Machinery}, \bibinfo{address}{New York, NY, USA}, Article \bibinfo{articleno}{82}, \bibinfo{numpages}{14}~pages.
\newblock
\showISBNx{9798400700958}
\href{https://doi.org/10.1145/3579371.3589350}{doi:\nolinkurl{10.1145/3579371.3589350}}


\bibitem[Jouppi et~al\mbox{.}(2017)]%
        {jouppi2017tpuv1}
\bibfield{author}{\bibinfo{person}{Norman~P. Jouppi}, \bibinfo{person}{Cliff Young}, \bibinfo{person}{Nishant Patil}, \bibinfo{person}{David Patterson}, \bibinfo{person}{Gaurav Agrawal}, \bibinfo{person}{Raminder Bajwa}, {et~al\mbox{.}}} \bibinfo{year}{2017}\natexlab{}.
\newblock \showarticletitle{In-Datacenter Performance Analysis of a Tensor Processing Unit}. In \bibinfo{booktitle}{\emph{Proceedings of the 44th Annual International Symposium on Computer Architecture}} (Toronto, ON, Canada) \emph{(\bibinfo{series}{ISCA'17})}. \bibinfo{publisher}{Association for Computing Machinery}, \bibinfo{address}{New York, NY, USA}, \bibinfo{pages}{1–12}.
\newblock
\showISBNx{9781450348928}
\href{https://doi.org/10.1145/3079856.3080246}{doi:\nolinkurl{10.1145/3079856.3080246}}


\bibitem[Kildall(1973)]%
        {kildall1973datflow}
\bibfield{author}{\bibinfo{person}{Gary~A. Kildall}.} \bibinfo{year}{1973}\natexlab{}.
\newblock \showarticletitle{A Unified Approach to Global Program Optimization}. In \bibinfo{booktitle}{\emph{Proceedings of the 1st Annual ACM SIGACT-SIGPLAN Symposium on Principles of Programming Languages}} (Boston, Massachusetts) \emph{(\bibinfo{series}{POPL'73})}. \bibinfo{publisher}{Association for Computing Machinery}, \bibinfo{address}{New York, NY, USA}, \bibinfo{pages}{194–206}.
\newblock
\showISBNx{9781450373494}
\href{https://doi.org/10.1145/512927.512945}{doi:\nolinkurl{10.1145/512927.512945}}


\bibitem[Laddad et~al\mbox{.}(2025)]%
        {laddad2025flo}
\bibfield{author}{\bibinfo{person}{Shadaj Laddad}, \bibinfo{person}{Alvin Cheung}, \bibinfo{person}{Joseph~M Hellerstein}, {and} \bibinfo{person}{Mae Milano}.} \bibinfo{year}{2025}\natexlab{}.
\newblock \showarticletitle{Flo: A Semantic Foundation for Progressive Stream Processing}.
\newblock \bibinfo{journal}{\emph{Proceedings of the ACM on Programming Languages}} \bibinfo{volume}{9}, \bibinfo{number}{POPL} (\bibinfo{year}{2025}), \bibinfo{pages}{241--270}.
\newblock


\bibitem[Lai et~al\mbox{.}(2019)]%
        {lai2019heterocl}
\bibfield{author}{\bibinfo{person}{Yi-Hsiang Lai}, \bibinfo{person}{Yuze Chi}, \bibinfo{person}{Yuwei Hu}, \bibinfo{person}{Jie Wang}, \bibinfo{person}{Cody~Hao Yu}, \bibinfo{person}{Yuan Zhou}, \bibinfo{person}{Jason Cong}, {and} \bibinfo{person}{Zhiru Zhang}.} \bibinfo{year}{2019}\natexlab{}.
\newblock \showarticletitle{HeteroCL: A Multi-Paradigm Programming Infrastructure for Software-Defined Reconfigurable Computing}. In \bibinfo{booktitle}{\emph{Proceedings of the 2019 ACM/SIGDA International Symposium on Field-Programmable Gate Arrays}} (Seaside, CA, USA) \emph{(\bibinfo{series}{FPGA'19})}. \bibinfo{publisher}{Association for Computing Machinery}, \bibinfo{address}{New York, NY, USA}, \bibinfo{pages}{242–251}.
\newblock
\showISBNx{9781450361378}
\href{https://doi.org/10.1145/3289602.3293910}{doi:\nolinkurl{10.1145/3289602.3293910}}


\bibitem[Lattner et~al\mbox{.}(2021)]%
        {chris2021mlir}
\bibfield{author}{\bibinfo{person}{Chris Lattner}, \bibinfo{person}{Mehdi Amini}, \bibinfo{person}{Uday Bondhugula}, \bibinfo{person}{Albert Cohen}, \bibinfo{person}{Andy Davis}, \bibinfo{person}{Jacques Pienaar}, \bibinfo{person}{River Riddle}, \bibinfo{person}{Tatiana Shpeisman}, \bibinfo{person}{Nicolas Vasilache}, {and} \bibinfo{person}{Oleksandr Zinenko}.} \bibinfo{year}{2021}\natexlab{}.
\newblock \showarticletitle{MLIR: Scaling Compiler Infrastructure for Domain Specific Computation}. In \bibinfo{booktitle}{\emph{Proceedings of the 2021 IEEE/ACM International Symposium on Code Generation and Optimization}} (Virtual Event, Republic of Korea) \emph{(\bibinfo{series}{CGO '21})}. \bibinfo{publisher}{IEEE Press}, \bibinfo{address}{Los Alamitos, CA, USA}, \bibinfo{pages}{2–14}.
\newblock
\showISBNx{9781728186139}
\href{https://doi.org/10.1109/CGO51591.2021.9370308}{doi:\nolinkurl{10.1109/CGO51591.2021.9370308}}


\bibitem[Li et~al\mbox{.}(2025)]%
        {li2025plaid}
\bibfield{author}{\bibinfo{person}{Zhaoying Li}, \bibinfo{person}{Pranav Dangi}, \bibinfo{person}{Chenyang Yin}, \bibinfo{person}{Thilini~Kaushalya Bandara}, \bibinfo{person}{Rohan Juneja}, \bibinfo{person}{Cheng Tan}, \bibinfo{person}{Zhenyu Bai}, {and} \bibinfo{person}{Tulika Mitra}.} \bibinfo{year}{2025}\natexlab{}.
\newblock \showarticletitle{Enhancing CGRA Efficiency Through Aligned Compute and Communication Provisioning}. In \bibinfo{booktitle}{\emph{Proceedings of the 30th ACM International Conference on Architectural Support for Programming Languages and Operating Systems, Volume 1}} (Rotterdam, Netherlands) \emph{(\bibinfo{series}{ASPLOS '25})}. \bibinfo{publisher}{Association for Computing Machinery}, \bibinfo{address}{New York, NY, USA}, \bibinfo{pages}{410–425}.
\newblock
\showISBNx{9798400706981}
\href{https://doi.org/10.1145/3669940.3707230}{doi:\nolinkurl{10.1145/3669940.3707230}}


\bibitem[NVIDIA(2017)]%
        {cutlass}
\bibfield{author}{\bibinfo{person}{NVIDIA}.} \bibinfo{year}{2017}\natexlab{}.
\newblock \bibinfo{title}{CUTLASS}.
\newblock \bibinfo{howpublished}{\url{https://github.com/NVIDIA/cutlass}}.
\newblock


\bibitem[NVIDIA(2022)]%
        {hopper}
\bibfield{author}{\bibinfo{person}{NVIDIA}.} \bibinfo{year}{2022}\natexlab{}.
\newblock \bibinfo{title}{NVIDIA Hopper Architecture}.
\newblock \bibinfo{howpublished}{\url{https://www.nvidia.com/en-us/data-center/technologies/hopper-architecture/}}.
\newblock


\bibitem[NVIDIA(2025a)]%
        {cute}
\bibfield{author}{\bibinfo{person}{NVIDIA}.} \bibinfo{year}{2025}\natexlab{a}.
\newblock \bibinfo{title}{CUTLASS Documentation: CuTe Layouts}.
\newblock
\urldef\tempurl%
\url{https://docs.nvidia.com/cutlass/media/docs/cpp/cute/01_layout.html}
\showURL{%
\tempurl}


\bibitem[NVIDIA(2025b)]%
        {blackwell}
\bibfield{author}{\bibinfo{person}{NVIDIA}.} \bibinfo{year}{2025}\natexlab{b}.
\newblock \bibinfo{title}{NVIDIA Blackwell Architecture Technical Brief}.
\newblock \bibinfo{howpublished}{\url{https://resources.nvidia.com/en-us-blackwell-architecture}}.
\newblock


\bibitem[Pal et~al\mbox{.}(2022)]%
        {debjit2022dac}
\bibfield{author}{\bibinfo{person}{Debjit Pal}, \bibinfo{person}{Yi-Hsiang Lai}, \bibinfo{person}{Shaojie Xiang}, \bibinfo{person}{Niansong Zhang}, \bibinfo{person}{Hongzheng Chen}, \bibinfo{person}{Jeremy Casas}, \bibinfo{person}{Pasquale Cocchini}, \bibinfo{person}{Zhenkun Yang}, \bibinfo{person}{Jin Yang}, \bibinfo{person}{Louis-No\"{e}l Pouchet}, {and} \bibinfo{person}{Zhiru Zhang}.} \bibinfo{year}{2022}\natexlab{}.
\newblock \showarticletitle{Accelerator Design with Decoupled Hardware Customizations: Benefits and Challenges: Invited}. In \bibinfo{booktitle}{\emph{Proceedings of the 59th ACM/IEEE Design Automation Conference}} (San Francisco, California) \emph{(\bibinfo{series}{DAC '22})}. \bibinfo{publisher}{Association for Computing Machinery}, \bibinfo{address}{New York, NY, USA}, \bibinfo{pages}{1351–1354}.
\newblock
\showISBNx{9781450391429}
\href{https://doi.org/10.1145/3489517.3530681}{doi:\nolinkurl{10.1145/3489517.3530681}}


\bibitem[Pope et~al\mbox{.}(2023)]%
        {pope2023scalellmtpu}
\bibfield{author}{\bibinfo{person}{Reiner Pope}, \bibinfo{person}{Sholto Douglas}, \bibinfo{person}{Aakanksha Chowdhery}, \bibinfo{person}{Jacob Devlin}, \bibinfo{person}{James Bradbury}, \bibinfo{person}{Jonathan Heek}, \bibinfo{person}{Kefan Xiao}, \bibinfo{person}{Shivani Agrawal}, {and} \bibinfo{person}{Jeff Dean}.} \bibinfo{year}{2023}\natexlab{}.
\newblock \showarticletitle{Efficiently scaling transformer inference}.
\newblock \bibinfo{journal}{\emph{Proceedings of machine learning and systems}}  \bibinfo{volume}{5} (\bibinfo{year}{2023}), \bibinfo{pages}{606--624}.
\newblock


\bibitem[Ragan-Kelley et~al\mbox{.}(2013)]%
        {jrk2013halide}
\bibfield{author}{\bibinfo{person}{Jonathan Ragan-Kelley}, \bibinfo{person}{Connelly Barnes}, \bibinfo{person}{Andrew Adams}, \bibinfo{person}{Sylvain Paris}, \bibinfo{person}{Fr\'{e}do Durand}, {and} \bibinfo{person}{Saman Amarasinghe}.} \bibinfo{year}{2013}\natexlab{}.
\newblock \showarticletitle{Halide: A Language and Compiler for Optimizing Parallelism, Locality, and Recomputation in Image Processing Pipelines}.
\newblock \bibinfo{journal}{\emph{SIGPLAN Not.}} \bibinfo{volume}{48}, \bibinfo{number}{6} (\bibinfo{date}{jun} \bibinfo{year}{2013}), \bibinfo{pages}{519–530}.
\newblock
\showISSN{0362-1340}
\href{https://doi.org/10.1145/2499370.2462176}{doi:\nolinkurl{10.1145/2499370.2462176}}


\bibitem[Rico et~al\mbox{.}(2024)]%
        {xdna_npu}
\bibfield{author}{\bibinfo{person}{Alejandro Rico}, \bibinfo{person}{Satyaprakash Pareek}, \bibinfo{person}{Javier Cabezas}, \bibinfo{person}{David Clarke}, \bibinfo{person}{Baris Ozgul}, \bibinfo{person}{Francisco Barat}, \bibinfo{person}{Yao Fu}, \bibinfo{person}{Stephan Münz}, \bibinfo{person}{Dylan Stuart}, \bibinfo{person}{Patrick Schlangen}, \bibinfo{person}{Pedro Duarte}, \bibinfo{person}{Sneha Date}, \bibinfo{person}{Indrani Paul}, \bibinfo{person}{Jian Weng}, \bibinfo{person}{Sonal Santan}, \bibinfo{person}{Vinod Kathail}, \bibinfo{person}{Ashish Sirasao}, {and} \bibinfo{person}{Juanjo Noguera}.} \bibinfo{year}{2024}\natexlab{}.
\newblock \showarticletitle{AMD XDNA NPU in Ryzen AI Processors}.
\newblock \bibinfo{journal}{\emph{IEEE Micro}} \bibinfo{volume}{44}, \bibinfo{number}{6} (\bibinfo{year}{2024}), \bibinfo{pages}{73--82}.
\newblock
\href{https://doi.org/10.1109/MM.2024.3423692}{doi:\nolinkurl{10.1109/MM.2024.3423692}}


\bibitem[Rioux and Zdancewic(2025)]%
        {nick2025parallelstreaming}
\bibfield{author}{\bibinfo{person}{Nick Rioux} {and} \bibinfo{person}{Steve Zdancewic}.} \bibinfo{year}{2025}\natexlab{}.
\newblock \showarticletitle{Functional Meaning for Parallel Streaming}.
\newblock \bibinfo{journal}{\emph{Proc. ACM Program. Lang.}} \bibinfo{volume}{9}, \bibinfo{number}{PLDI}, Article \bibinfo{articleno}{196} (\bibinfo{date}{June} \bibinfo{year}{2025}), \bibinfo{numpages}{25}~pages.
\newblock
\href{https://doi.org/10.1145/3729299}{doi:\nolinkurl{10.1145/3729299}}


\bibitem[Rucker et~al\mbox{.}(2024)]%
        {rucker2024revet}
\bibfield{author}{\bibinfo{person}{Alexander~C. Rucker}, \bibinfo{person}{Shiv Sundram}, \bibinfo{person}{Coleman Smith}, \bibinfo{person}{Matthew Vilim}, \bibinfo{person}{Raghu Prabhakar}, \bibinfo{person}{Fredrik Kjølstad}, {and} \bibinfo{person}{Kunle Olukotun}.} \bibinfo{year}{2024}\natexlab{}.
\newblock \showarticletitle{Revet: A Language and Compiler for Dataflow Threads}. In \bibinfo{booktitle}{\emph{2024 IEEE International Symposium on High-Performance Computer Architecture (HPCA)}}. \bibinfo{publisher}{IEEE Computer Society}, \bibinfo{address}{Los Alamitos, CA, USA}, \bibinfo{pages}{1--14}.
\newblock
\href{https://doi.org/10.1109/HPCA57654.2024.00016}{doi:\nolinkurl{10.1109/HPCA57654.2024.00016}}


\bibitem[Sabne(2020)]%
        {xla}
\bibfield{author}{\bibinfo{person}{Amit Sabne}.} \bibinfo{year}{2020}\natexlab{}.
\newblock \bibinfo{title}{XLA : Compiling Machine Learning for Peak Performance}.
\newblock


\bibitem[Tenstorrent(2024)]%
        {tenstorrent}
\bibfield{author}{\bibinfo{person}{Tenstorrent}.} \bibinfo{year}{2024}\natexlab{}.
\newblock \bibinfo{title}{Blackhole}.
\newblock \bibinfo{howpublished}{\url{https://tenstorrent.com/en/hardware/blackhole}}.
\newblock


\bibitem[Thomas et~al\mbox{.}(2020)]%
        {thomas2020fleet}
\bibfield{author}{\bibinfo{person}{James Thomas}, \bibinfo{person}{Pat Hanrahan}, {and} \bibinfo{person}{Matei Zaharia}.} \bibinfo{year}{2020}\natexlab{}.
\newblock \showarticletitle{Fleet: A Framework for Massively Parallel Streaming on FPGAs}. In \bibinfo{booktitle}{\emph{Proceedings of the Twenty-Fifth International Conference on Architectural Support for Programming Languages and Operating Systems}} (Lausanne, Switzerland) \emph{(\bibinfo{series}{ASPLOS '20})}. \bibinfo{publisher}{Association for Computing Machinery}, \bibinfo{address}{New York, NY, USA}, \bibinfo{pages}{639–651}.
\newblock
\showISBNx{9781450371025}
\href{https://doi.org/10.1145/3373376.3378495}{doi:\nolinkurl{10.1145/3373376.3378495}}


\bibitem[Tillet et~al\mbox{.}(2019)]%
        {tillet2019triton}
\bibfield{author}{\bibinfo{person}{Philippe Tillet}, \bibinfo{person}{H.~T. Kung}, {and} \bibinfo{person}{David Cox}.} \bibinfo{year}{2019}\natexlab{}.
\newblock \showarticletitle{Triton: An Intermediate Language and Compiler for Tiled Neural Network Computations}. In \bibinfo{booktitle}{\emph{Proceedings of the 3rd ACM SIGPLAN International Workshop on Machine Learning and Programming Languages}}. \bibinfo{publisher}{ACM}, \bibinfo{address}{New York, NY, USA}, \bibinfo{pages}{10--19}.
\newblock


\bibitem[Vaswani et~al\mbox{.}(2017)]%
        {vaswani2017transformer}
\bibfield{author}{\bibinfo{person}{Ashish Vaswani}, \bibinfo{person}{Noam Shazeer}, \bibinfo{person}{Niki Parmar}, \bibinfo{person}{Jakob Uszkoreit}, \bibinfo{person}{Llion Jones}, \bibinfo{person}{Aidan~N. Gomez}, \bibinfo{person}{\L{}ukasz Kaiser}, {and} \bibinfo{person}{Illia Polosukhin}.} \bibinfo{year}{2017}\natexlab{}.
\newblock \showarticletitle{Attention is All You Need}. In \bibinfo{booktitle}{\emph{Proceedings of the 31st International Conference on Neural Information Processing Systems}} (Long Beach, California, USA) \emph{(\bibinfo{series}{NIPS'17})}. \bibinfo{publisher}{Curran Associates Inc.}, \bibinfo{address}{Red Hook, NY, USA}, \bibinfo{pages}{6000–6010}.
\newblock
\showISBNx{9781510860964}


\bibitem[Wang et~al\mbox{.}(2025)]%
        {wang2025tilelang}
\bibfield{author}{\bibinfo{person}{Lei Wang}, \bibinfo{person}{Yu Cheng}, \bibinfo{person}{Yining Shi}, \bibinfo{person}{Zhengju Tang}, \bibinfo{person}{Zhiwen Mo}, \bibinfo{person}{Wenhao Xie}, \bibinfo{person}{Lingxiao Ma}, \bibinfo{person}{Yuqing Xia}, \bibinfo{person}{Jilong Xue}, \bibinfo{person}{Fan Yang}, {and} \bibinfo{person}{Zhi Yang}.} \bibinfo{year}{2025}\natexlab{}.
\newblock \bibinfo{title}{TileLang: A Composable Tiled Programming Model for AI Systems}.
\newblock
\showeprint[arxiv]{2504.17577}~[cs.LG]
\urldef\tempurl%
\url{https://arxiv.org/abs/2504.17577}
\showURL{%
\tempurl}


\bibitem[Xiang et~al\mbox{.}(2022)]%
        {xiang2022heteroflow}
\bibfield{author}{\bibinfo{person}{Shaojie Xiang}, \bibinfo{person}{Yi-Hsiang Lai}, \bibinfo{person}{Yuan Zhou}, \bibinfo{person}{Hongzheng Chen}, \bibinfo{person}{Niansong Zhang}, \bibinfo{person}{Debjit Pal}, {and} \bibinfo{person}{Zhiru Zhang}.} \bibinfo{year}{2022}\natexlab{}.
\newblock \showarticletitle{HeteroFlow: An Accelerator Programming Model with Decoupled Data Placement for Software-Defined FPGAs}. In \bibinfo{booktitle}{\emph{Proceedings of the 2022 ACM/SIGDA International Symposium on Field-Programmable Gate Arrays}} (Virtual Event, USA) \emph{(\bibinfo{series}{FPGA'22})}. \bibinfo{publisher}{Association for Computing Machinery}, \bibinfo{address}{New York, NY, USA}, \bibinfo{pages}{78–88}.
\newblock
\showISBNx{9781450391498}
\href{https://doi.org/10.1145/3490422.3502369}{doi:\nolinkurl{10.1145/3490422.3502369}}


\bibitem[Xie et~al\mbox{.}(2022)]%
        {xie2022p2}
\bibfield{author}{\bibinfo{person}{Ningning Xie}, \bibinfo{person}{Tamara Norman}, \bibinfo{person}{Dominik Grewe}, {and} \bibinfo{person}{Dimitrios Vytiniotis}.} \bibinfo{year}{2022}\natexlab{}.
\newblock \showarticletitle{Synthesizing optimal parallelism placement and reduction strategies on hierarchical systems for deep learning}.
\newblock \bibinfo{journal}{\emph{Proceedings of Machine Learning and Systems}}  \bibinfo{volume}{4} (\bibinfo{year}{2022}), \bibinfo{pages}{548--566}.
\newblock


\bibitem[Xilinx(2025)]%
        {mliraie}
\bibfield{author}{\bibinfo{person}{Xilinx}.} \bibinfo{year}{2025}\natexlab{}.
\newblock \bibinfo{title}{MLIR-AIE Dialect}.
\newblock \bibinfo{howpublished}{\url{https://github.com/Xilinx/mlir-aie}}.
\newblock


\bibitem[Xilinx(2022a)]%
        {xilinxAIE}
\bibfield{author}{\bibinfo{person}{AMD Xilinx}.} \bibinfo{year}{2022}\natexlab{a}.
\newblock \bibinfo{title}{{AI Engines and Their Applications}}.
\newblock
\urldef\tempurl%
\url{https://www.xilinx.com/content/dam/xilinx/support/documents/white_papers/wp506-ai-engine.pdf}
\showURL{%
\tempurl}


\bibitem[Xilinx(2022b)]%
        {vitis_hls_library}
\bibfield{author}{\bibinfo{person}{AMD Xilinx}.} \bibinfo{year}{2022}\natexlab{b}.
\newblock \bibinfo{title}{Vitis Accelerated Libraries}.
\newblock \bibinfo{howpublished}{\url{https://github.com/Xilinx/Vitis_Libraries}}.
\newblock


\bibitem[Xilinx(2022c)]%
        {vitis_ai}
\bibfield{author}{\bibinfo{person}{AMD Xilinx}.} \bibinfo{year}{2022}\natexlab{c}.
\newblock \bibinfo{title}{Vitis AI: Adaptable \& Real-Time AI Inference Acceleration}.
\newblock \bibinfo{howpublished}{\url{https://github.com/Xilinx/Vitis-AI}}.
\newblock


\bibitem[Xilinx(2023)]%
        {vitis_hls_2023}
\bibfield{author}{\bibinfo{person}{AMD Xilinx}.} \bibinfo{year}{2023}\natexlab{}.
\newblock \bibinfo{title}{Vitis HLS v2023.2}.
\newblock \bibinfo{howpublished}{\url{https://www.xilinx.com/products/design-tools/vitis/vitis-platform.html}}.
\newblock


\bibitem[Xu et~al\mbox{.}(2021)]%
        {xu2021gspmd}
\bibfield{author}{\bibinfo{person}{Yuanzhong Xu}, \bibinfo{person}{HyoukJoong Lee}, \bibinfo{person}{Dehao Chen}, \bibinfo{person}{Blake Hechtman}, \bibinfo{person}{Yanping Huang}, \bibinfo{person}{Rahul Joshi}, \bibinfo{person}{Maxim Krikun}, \bibinfo{person}{Dmitry Lepikhin}, \bibinfo{person}{Andy Ly}, \bibinfo{person}{Marcello Maggioni}, \bibinfo{person}{Ruoming Pang}, \bibinfo{person}{Noam Shazeer}, \bibinfo{person}{Shibo Wang}, \bibinfo{person}{Tao Wang}, \bibinfo{person}{Yonghui Wu}, {and} \bibinfo{person}{Zhifeng Chen}.} \bibinfo{year}{2021}\natexlab{}.
\newblock \bibinfo{title}{GSPMD: General and Scalable Parallelization for ML Computation Graphs}.
\newblock
\showeprint[arxiv]{2105.04663}~[cs.DC]
\urldef\tempurl%
\url{https://arxiv.org/abs/2105.04663}
\showURL{%
\tempurl}


\bibitem[Ye et~al\mbox{.}(2024)]%
        {ye2024hida}
\bibfield{author}{\bibinfo{person}{Hanchen Ye}, \bibinfo{person}{Hyegang Jun}, {and} \bibinfo{person}{Deming Chen}.} \bibinfo{year}{2024}\natexlab{}.
\newblock \showarticletitle{Hida: A hierarchical dataflow compiler for high-level synthesis}. In \bibinfo{booktitle}{\emph{Proceedings of the 29th ACM International Conference on Architectural Support for Programming Languages and Operating Systems, Volume 1}}. \bibinfo{publisher}{ACM}, \bibinfo{address}{New York, NY, USA}, \bibinfo{pages}{215--230}.
\newblock


\bibitem[Zheng et~al\mbox{.}(2022)]%
        {zheng2022alpa}
\bibfield{author}{\bibinfo{person}{Lianmin Zheng}, \bibinfo{person}{Zhuohan Li}, \bibinfo{person}{Hao Zhang}, \bibinfo{person}{Yonghao Zhuang}, \bibinfo{person}{Zhifeng Chen}, \bibinfo{person}{Yanping Huang}, \bibinfo{person}{Yida Wang}, \bibinfo{person}{Yuanzhong Xu}, \bibinfo{person}{Danyang Zhuo}, \bibinfo{person}{Eric~P. Xing}, \bibinfo{person}{Joseph~E. Gonzalez}, {and} \bibinfo{person}{Ion Stoica}.} \bibinfo{year}{2022}\natexlab{}.
\newblock \showarticletitle{Alpa: Automating Inter- and {Intra-Operator} Parallelism for Distributed Deep Learning}. In \bibinfo{booktitle}{\emph{16th USENIX Symposium on Operating Systems Design and Implementation (OSDI 22)}}. \bibinfo{publisher}{USENIX Association}, \bibinfo{address}{Carlsbad, CA}, \bibinfo{pages}{559--578}.
\newblock
\showISBNx{978-1-939133-28-1}
\urldef\tempurl%
\url{https://www.usenix.org/conference/osdi22/presentation/zheng-lianmin}
\showURL{%
\tempurl}


\bibitem[Zhou et~al\mbox{.}(2025)]%
        {zhou2025linearlayout}
\bibfield{author}{\bibinfo{person}{Keren Zhou}, \bibinfo{person}{Mario Lezcano}, \bibinfo{person}{Adam Goucher}, \bibinfo{person}{Akhmed Rakhmati}, \bibinfo{person}{Jeff Niu}, \bibinfo{person}{Justin Lebar}, \bibinfo{person}{Pawel Szczerbuk}, \bibinfo{person}{Peter Bell}, \bibinfo{person}{Phil Tillet}, \bibinfo{person}{Thomas Raoux}, {et~al\mbox{.}}} \bibinfo{year}{2025}\natexlab{}.
\newblock \bibinfo{title}{Linear Layouts: Robust Code Generation of Efficient Tensor Computation Using $\mathbb{F}_2$}.
\newblock
\showeprint[arxiv]{2505.23819}~[cs.PL]
\urldef\tempurl%
\url{https://arxiv.org/abs/2505.23819}
\showURL{%
\tempurl}


\bibitem[Zhuang et~al\mbox{.}(2025)]%
        {zhuang2025aries}
\bibfield{author}{\bibinfo{person}{Jinming Zhuang}, \bibinfo{person}{Shaojie Xiang}, \bibinfo{person}{Hongzheng Chen}, \bibinfo{person}{Niansong Zhang}, \bibinfo{person}{Zhuoping Yang}, \bibinfo{person}{Tony Mao}, \bibinfo{person}{Zhiru Zhang}, {and} \bibinfo{person}{Peipei Zhou}.} \bibinfo{year}{2025}\natexlab{}.
\newblock \showarticletitle{ARIES: An Agile MLIR-Based Compilation Flow for Reconfigurable Devices with AI Engines}. In \bibinfo{booktitle}{\emph{Proceedings of the 2025 ACM/SIGDA International Symposium on Field Programmable Gate Arrays}}. \bibinfo{publisher}{ACM}, \bibinfo{address}{New York, NY, USA}, \bibinfo{pages}{92--102}.
\newblock


\end{thebibliography}

\newpage
\appendix
\section{Daisy-Chain Systolic Array}
\label{appendix:sa}
The \Name implementation of the daisy-chain systolic array is shown below, where \mintinline{python}{dato.meta_if} and \mintinline{python}{dato.meta_else} are compile-time conditional branches that incur no hardware overhead.

\begin{minted}[linenos,
               fontsize=\scriptsize,
               xleftmargin=1.8em,
               escapeinside=||,
               autogobble]{python}
import dato
from dato.ir.types import int8, UInt, Stream
import dato.dsl as dsl

M, N, K = 64, 64, 64
Rt, Ct = 16, 16

P0, P1 = Rt + 2, Ct + 2

def top():
    L3_A: Stream[UInt(Rt * 8), 4]
    L3_B: Stream[UInt(Ct * 8), 4]
    L3_C: Stream[UInt(Rt * 8), 4]

    L2_A: Stream[UInt(Rt * 8)][P0 - 1]
    L2_B: Stream[UInt(Ct * 8)][P1 - 1]

    L1_C: Stream[UInt(Rt * 8)][Rt, Ct]
    L2_C: Stream[UInt(Rt * 8)][Ct]

    fifo_A: Stream[int8][Rt, Ct]
    fifo_B: Stream[int8][Rt, Ct]

    @dato.task(mapping=[1])
    def offchip_loadA(A_Packed: UInt(Rt * 8)[M * K // Rt]):
        for mt, nt in dsl.grid(M // Rt, N // Ct):
            for k in range(K):
                L3_A.put(A_Packed[mt * K + k])

    @dato.task(mapping=[1])
    def offchip_loadB(B_Packed: UInt(Ct * 8)[K * N // Ct]):
        for mt, nt in dsl.grid(M // Rt, N // Ct):
            for k in range(K):
                L3_B.put(B_Packed[nt * K + k])

    @dato.task(mapping=[P0, P1])
    def gemm_systolic():
        i, j = dato.get_tid()
        # peripheral kernels
        with dato.meta_if(i == 0 and j == 0):
            for mt, nt in dsl.grid(M // Rt, N // Ct):
                for k in range(K):
                    L2_A[1].put(L3_A.get())
                    L2_B[1].put(L3_B.get())

        with dato.meta_elif(i == P0 - 1 and j == P1 - 1):
            for mt, nt in dsl.grid(M // Rt, N // Ct):
                for n in range(Ct):
                    L3_C.put(L2_C[Ct - 1].get())

        with dato.meta_elif(i in {0, P0 - 1} and j in {0, P1 - 1}):
            pass

        with dato.meta_elif(j == 0):
            # i > 0, the first column
            for mt, nt in dsl.grid(M // Rt, N // Ct):
                for k in range(K):
                    a = L2_A[i].get()
                    # unpack data
                    fifo_A[i - 1, 0].put(a[8 * (i - 1) : 8 * i])
                    with dato.meta_if(i < Rt):
                        L2_A[i + 1].put(a)

        with dato.meta_elif(i == 0):
            # j > 0, the first row
            for mt, nt in dsl.grid(M // Rt, N // Ct):
                for k in range(K):
                    b = L2_B[j].get()
                    fifo_B[0, j - 1].put(b[8 * (j - 1) : 8 * j])
                    with dato.meta_if(j < Ct):
                        L2_B[j + 1].put(b)

        with dato.meta_elif(i == P0 - 1):
            for mt, nt in dsl.grid(M // Rt, N // Ct):
                c_C = L1_C[i - 2, Ct - j].get()
                L2_C[j - 1].put(c_C)
                with dato.meta_if(j != 1):
                    for ind in range(j - 1):
                        L2_C[j - 1].put(L2_C[j - 2].get())

        with dato.meta_elif(j == P1 - 1):
            pass

        # main body
        with dato.meta_else():
            for mt, nt in dsl.grid(M // Rt, N // Ct):
                c: int8 = 0
                for k in range(K):
                    a: int8 = fifo_A[i - 1, j - 1].get()
                    b: int8 = fifo_B[i - 1, j - 1].get()
                    c += a * b
                    with dato.meta_if(j < Ct):
                        fifo_A[i - 1, j].put(a)
                    with dato.meta_if(i < Rt):
                        fifo_B[i, j - 1].put(b)

                with dato.meta_if(i == 1):
                    packed_tmp: UInt(M * 8) = 0
                with dato.meta_else():
                    packed_tmp: UInt(M * 8) = L1_C[i - 2, j - 1].get()

                packed_c: UInt(M * 8) = 0
                for m in range(Rt):
                    if m == i - 1:
                        packed_c[m * 8 : (m + 1) * 8] = c
                    else:
                        packed_c[m * 16 : (m + 1) * 16] = packed_tmp[
                            m * 16 : (m + 1) * 16
                        ]
                L1_C[i - 1, j - 1].put(packed_c)

    @dato.task(mapping=[1])
    def offchip_store(C_Packed: UInt(Rt * 8)[M * N // Rt]):
        for mt, nt in dsl.grid(M // Rt, N // Ct):
            for n in range(Ct):
                C_Packed[mt * N + nt * Ct + n] = L3_C.get()
\end{minted}

\end{document}